**Hosting is not capturing: an exposure–capability framework and Hungary's electric-vehicle battery bet**

Zoltán Bartha, zoli@ekon.me

https://orcid.org/0000-0002-9208-7881

**Abstract**

How should a dependent economy respond when a polycrisis turns economic integration from a source of efficiency into a surface of vulnerability? This paper develops the exposure–capability framework, which maps a country's strategic options along two dimensions — the concentration of its external dependence and the depth of its productive capability — and yields four ideal strategies. It applies the framework to Hungary's large-scale bet on electric-vehicle battery manufacturing, asking what the concentrated exposure purchased: movement toward a higher position in the battery value chain, as a catch-up gamble, or the reproduction of the country's existing position under a new product, as dependent entrenchment. Using a firm-level panel of Hungary's largest companies for 2010–2024, it tests four hypotheses and examines the conditions attached to the support that financed the investments. The sector has expanded enormously, but its value capture neither rises beyond the recovery from construction losses nor shifts up the chain: production remains concentrated in the lowest-value stage, matures only to the level of comparably young foreign-owned firms, and does not cover its own costs. The support that financed it required investment volume, revenue and employment retention; capability-relevant criteria entered the decree as considerations rather than conditions, and a capability-contingent instrument within the same scheme went unused. On the evidence to 2024 the bet is a lateral substitution at constant quality, in which battery-cell fabrication replaces automotive assembly at the same dependent position. Hosting the green transition's supply chain, the case shows, is not the same as capturing its value.

## 1. Introduction

Hungary's economic strategy has for several years combined a rhetoric of national sovereignty with the deepening of external dependence. In 2022 the Chinese battery manufacturer CATL announced a €7.3 billion gigafactory in Debrecen, the largest foreign investment in the country's history (Transport & Environment, 2025). It was the most conspicuous move in a strategy that has made a country of fewer than ten million people one of Europe's principal sites of electric-vehicle battery production, drawing Korean and Chinese capital on a scale out of proportion to its size (Czirfusz, 2023). These investments sit atop an economy already built on foreign capital: a manufacturing base owned and directed largely from abroad, assembling cars and components for export (Nölke & Vliegenthart, 2009; Pavlinek et al., 2017). They arrived alongside a deepening rather than a loosening of energy dependence on Russia, preserved through EU exemptions even as the rest of the bloc moved to sever those ties (Malak et al., 2025). The government that pursued them campaigned on sovereignty and on an Eastern Opening intended to lessen Western dependence.

The largest of these investments illustrates the tension. CATL is the kind of producer the European Union has charged with competing on the strength of unfair state subsidies, the finding behind the countervailing duties the Commission imposed on Chinese electric vehicles in late 2024 (European Commission, 2024b). Yet its Debrecen plant is itself underwritten by some €800 million in Hungarian grants, tax allowances, and dedicated infrastructure (Tagliapietra et al., 2025).

Such an arrangement may be read as inconsistency. It can also be read as strategy, and the second reading is the one pursued here. The environment in which states make economic policy has changed: the integration that once promised efficiency now also exposes economies to disruption and coercion, as supply chains, energy links, and technology flows are turned into instruments of pressure. In such a world no economy holds a neutral position, and each must decide how to manage its dependence. For a small, dependent economy the range of available responses is narrow and the stakes are high. Hungary's battery bet is one such response — a wager that hosting the supply chain of the green transition will convert dependence into advantage. This paper asks what the wager purchased in the terms that matter for the country's long-run position: has the bet moved the targeted firms toward a higher position in the battery value chain — a catch-up gamble — or has it reproduced the position they entered at, under a new product — dependent entrenchment?

The two outcomes are easily mistaken for one another. Both open with a large industrial build-out — new plants, jobs, exports — and on the surface the battery bet has plainly delivered one. But scale alone cannot tell upgrading from entrenchment: a sector can grow enormously while capturing little of the value that passes through it, sustained by subsidy rather than by what it makes. It also means having a benchmark: a clear account of the strategies that were available, against which the bet's position can be judged. The difficulty is analytical and empirical at once: every dependent economy weighing a similar bet faces the same problem of knowing whether it is climbing or running in place.

This paper meets that problem in two steps. First, it develops a general framework — the exposure–capability framework — that maps the strategic responses open to a dependent economy along the two dimensions the changed environment has made decisive: the shape of its external exposure, concentrated or diversified, and the depth of its productive capability, rent-extractive or capacity-building. The framework yields four ideal strategies and locates the bet among them, supplying the benchmark against which it can be assessed. Second, it tests the bet empirically, against that benchmark, using a firm-level panel of Hungary's largest companies over 2010–2024 that isolates the battery firms and sets them beside the established foreign-owned export base and the sheltered domestic economy. The test separates the volume the bet generated from the position it purchased, and traces whether its returns are carried by the firms' own productivity or by the state. The finding, reached across four lines of evidence, is that the battery bet is dependent entrenchment rather than a catch-up gamble. The bet has placed its firms at the lowest-value stage of the battery chain, where their value capture matures only to the level of the export model they join, sustained by discretionary subsidy at an intensity no other part of the economy receives. It is a lateral substitution at constant quality: battery fabrication put in the place of automotive assembly, at the same dependent position, and at the same level of value capture.

The paper proceeds as follows. Section 2 reconstructs the changed environment and derives the two dimensions on which national strategy now turns. Section 3 builds these into the exposure–capability framework and its four strategies. Section 4 locates the bet within the framework and sets out the hypotheses that separate a catch-up gamble from dependent entrenchment. Section 5 presents the firm-level panel and tests them. Section 6 concludes.

## 2. Diagnosis and strategy in a polycrisis

National economic strategies rest on a prior act of interpretation. Before a government chooses how to position its economy, it must read the environment that economy inhabits — what that environment rewards and punishes, where it offers opportunity and where it conceals risk. Strategy, in this sense, is downstream of diagnosis (Rumelt, 2017): policies follow from a reading of the world, and when that reading changes, strategies change with it. For much of the era of hyperglobalisation (Rodrik, 2012) the prevailing diagnosis was settled enough to need little restating — an open world economy in which integration brought efficiency and growth (Subramanian & Kessler, 2014), and a country's task was to find and occupy its place within it. That diagnosis no longer holds. The environment that national strategies must now read is one of compounding disorder, in which economic integration is as much a source of vulnerability as of gain (Keohane & Nye, 2012), and the question facing every state is no longer simply how to integrate but how to position itself against the risks integration carries.

This section reconstructs that changed environment and the strategic problem it poses, in three steps. It begins with the polycrisis, understood as an analytical tool: a way of grasping how distinct crises, interacting, transmit their shocks through the channels of economic interdependence and so turn a country's external relationships into a surface of vulnerability (Lawrence et al., 2024; Tooze, 2022). It then traces how the components of that polycrisis have moved the international order from one governed by rules to one governed by power, in which interdependence becomes a source of leverage rather than only of mutual benefit (Farrell & Newman, 2019). And it argues that this shift makes strategy unavoidable: where the market no longer guarantees security, every state must actively manage its position, and the dependent economies of the semi-periphery must do so from the weakest hand. From this account two dimensions emerge — the shape of a country's external exposure and the depth of its productive capability — which the following section develops into the framework.

### 2.1 The polycrisis as analytical tool

I take the polycrisis not as a synonym for hard times but as a specific claim about how crises relate. The term, coined by Morin and Kern (1999) and revived in European policy debate before it reached wider currency around 2022, has acquired a substantive definition in recent work: a global polycrisis is the causal entanglement of crises across multiple systems, such that their combined harm exceeds the sum of its parts (Lawrence et al., 2024; Tooze, 2022). The emphasis falls on entanglement. What distinguishes a polycrisis from a mere accumulation of troubles is that the crises interact — feeding, amplifying, and reshaping one another — so that none can be understood, or addressed, in isolation.

In the account I draw on, a crisis begins when a fast-moving trigger meets a slow-building stress and pushes a system out of its working equilibrium; crises become a polycrisis when these disturbances travel between systems, through shared stresses, knock-on effects, and feedbacks that loop back to worsen the original disturbance (Lawrence et al., 2024). The mechanism is one of transmission: the harm of any single shock is set by how tightly its host system is coupled to others, and by the channels along which a disturbance in one can propagate into the rest.

For an economy, those channels are the relationships through which it is integrated into the world — its supply chains, its energy imports, its financing, its access to technology and markets. This is the move that makes the polycrisis bear directly on national economic strategy. The same connections that integration builds, and that deliver its gains, are also the conduits along which compounding shocks arrive: a pandemic becomes a supply shock, a war becomes an energy shock, a rivalry becomes a technology shock, each travelling into a national economy through the very links that bind it to the world. That the polycrisis bears specifically on economic globalisation, and not only on global systems in the abstract, is the theme Helleiner (2024) develops. Integration, on this reading, is double-edged by construction. It is a source of efficiency and a surface of vulnerability at once — the insight the older interdependence literature already carried in its distinction between benign sensitivity and the costlier vulnerability that asymmetric dependence creates (Hirschman, 1945; Keohane & Nye, 2012), now generalised by the polycrisis from a bilateral relationship to a whole interconnected system.

A diagnosis framed as exposure raises the two questions on which a country's predicament turns. The first concerns the shape of its exposure — whether its critical dependencies are concentrated on a few partners and chokepoints, where a single disruption can be decisive, or spread widely enough to absorb a shock. The second concerns the depth of its capability — whether it commands the productive capacity to adapt, substitute, and reposition when a channel is disrupted or weaponised, or must simply endure. These two properties are what the polycrisis foregrounds as the determinants of how a country fares in a disordered world.

## 2.2 From rules to power

The order that the hyperglobalisation diagnosis took for granted rested on a political bargain: interdependence was treated as mutual gain, trade disputes were settled by rules through multilateral institutions, production was allocated by comparative advantage, and economics was kept, in principle, separate from security (Rodrik, 2012). The components of the polycrisis dismantled that bargain by revealing what the integrated system had become. The pandemic exposed concentrated supply chains as points of fragility. Russia's manipulation of gas supply turned a routine commercial dependency into an instrument of coercion (Malak et al., 2025). The contest between the United States and China made semiconductors and advanced technology objects of export control rather than ordinary trade, each side racing to secure the chokepoints the other might close (Miller, 2022). And the green transition, even as it promised to resolve one crisis, manufactured new dependencies — on the critical minerals and battery supply chains that decarbonisation requires, whose processing is concentrated in a handful of states, China above all (Bordoff & O'Sullivan, 2022).

The common thread is that integration's channels had become channels of leverage. This is the logic of weaponised interdependence: where economic networks pass through chokepoints, whoever controls those points can turn access into a tool of coercion, converting the connections that bind economies together into means of pressure (Farrell & Newman, 2019). What the hyperglobalisation order read as efficient division of labour, the polycrisis re-reads as a map of vulnerabilities — and, for those holding the chokepoints, of power.

The result is an order in which economic relations are governed increasingly by strategic power rather than by rules. The line between economic and security policy, central to the old bargain, has dissolved into a single domain of economic statecraft (Blackwill & Harris, 2016). States have returned to the industrial policy the previous consensus disdained, now reframed as a matter of resilience and security rather than of growth alone (Juhász et al., 2024). This is the environment that national strategies must now read: not a market that rewards integration as such, but a field of power in which every economic relationship carries a strategic charge.

## 2.3 The strategic imperative

If the environment is one of power rather than rules, no economy can occupy a neutral position within it. Under the polycrisis diagnosis the market no longer guarantees security, because the channels of integration are also the channels of coercion. A state must therefore take its position deliberately, managing the shape of its exposure and the depth of its capability rather than leaving them to be set by others. In such an order the absence of a strategy is not neutrality but drift: to leave one's dependencies unmanaged is to leave them available for another to exploit.

The new diagnosis has already reshaped policy. The great powers moved first and hardest, each in the idiom of its position: the United States through subsidy and reshoring, the European Union through the language of strategic autonomy and a net-zero industrial agenda, China through 'dual circulation' and the pursuit of self-reliance where it remains exposed. The specifics differ, but the move is the same: the deliberate management of exposure and the building of capability where dependence has turned dangerous (Hauge et al., 2025).

The imperative is general but unevenly met, and it falls hardest on the dependent economies of the semi-periphery, which face it from the weakest hand. Their exposure is high, because integration is deep and their critical dependencies concentrated; their capability is thin, because the productive capacity that would let them adapt is largely foreign-owned and externally directed (Bohle & Greskovits, 2012; Nölke & Vliegenthart, 2009; Ricz & Éltető, 2025). For these states the strategic question is sharper and the room for manoeuvre narrower: they cannot build capability at will, nor easily diversify dependencies that are structural. What they can choose is which new dependencies to take on, and on what terms — and whether the capability that arrives with foreign investment is genuinely acquired or merely hosted. Hungary, woven deep into the European automotive complex and now staking its industrial future on batteries, is an exemplary case of a semi-peripheral economy making this kind of wager.

**3. The exposure-capability framework**

Two claims from the preceding section carry forward: the environment is a polycrisis of interacting system transitions that no actor controls (Lawrence et al., 2024), and weaponised interdependence concentrates power in the actors controlling the network's critical nodes (Farrell & Newman, 2019). There is therefore no universally rational response, because leverage is a function of position. I make that claim tractable here by developing a typology.

**3.1 The two dimensions of response strategies**

A response to weaponised interdependence must resolve two questions that do not reduce to one another. The first concerns exposure: for each resource on which the economy depends — energy, finance, technology, inputs, market access — through how many independent channels can it be obtained, and how much of that access runs through channels a single external actor controls? The second concerns development: is the economy's growth driven by rising productive capability — the upgrading that yields long-run efficiency gains — or is its output merely propped up behind barriers to competition, while capability stays flat? The questions are independent — an economy may diversify its dependencies yet remain in low-value activity, or climb the technological ladder yet rely on a single supplier — so two dimensions are required.

**3.1.1 Axis A: dependency structure (concentration ←→ diversification)**

The first axis derives from weaponised interdependence. Farrell & Newman (2019) show that economic networks contain hubs through which a disproportionate share of flows must pass, and that states with jurisdiction over those hubs can convert them into coercion, by extracting information (the 'panopticon effect') or denying access (the 'chokepoint effect'). In the polycrisis framework's own terms, power is the capacity to manipulate the vectors — money, energy, information — that move through the system's conduits (Lawrence et al., 2024, following Dahl, 1957), and weaponised interdependence is the capture of those conduits: dollar clearing, semiconductor supply chains, pipelines, final-goods markets, and so on. Exposure is therefore a function of concentration: where a critical vector runs through a single foreign-controlled conduit, the controller holds a chokepoint; where it can be obtained through several, none can. Diversification is the structural defence, concentration the structural vulnerability — though, as the resilience literature notes, concentration is also efficiency-maximising, so diversification rebuilds resilience at the price of redundancy and efficiency (Folke et al., 2010; Walker et al., 2004).

Movement toward diversification is contested: a hub-holder's leverage depends on its target's continued concentration, so hub-holders resist their dependents' diversification. The small-state hedging literature offers independent support: Kuik (2021) notes that great powers actively dislike hedging, so the room available to small states to hedge is shrinking.

**3.1.2 Axis B: developmental content (rent-extraction ←→ capacity-building)**

The second axis runs from rent-extraction to capacity-building. At the rent-extraction pole, gains are secured not by creating value but by capturing it: elites use political power and influence — regulatory privileges and political–business networks — to engineer the scarcities from which rents flow, an activity that is "directly unproductive" (Bhagwati, 1982), reallocates resources away from productive uses (Congleton et al., 2008; Krueger, 1974; Tullock, 1967), and so erodes efficiency and long-run growth. Its deeper cost is dynamic: a firm that secures its returns by entrenching itself against competition has little reason to search for better methods, so rent-extraction suppresses the innovation on which long-run growth depends (Murphy et al., 1993). Bartha (2026) documents this across

Hungarian agriculture, construction, and finance in the 2010s, where politically favoured firms captured rising profit shares with no matching productivity gain.

The opposing pole, capacity-building, instead builds autonomous productive capability — committing resources to upgrading rather than to capturing rents. This marks the critical transition from merely expanding routine production capacity to accumulating dynamic technological capability (Bell & Pavitt, 1993). The two poles differ in what an economy holds when the investment matures: capacity-building leaves behind productive knowledge that outlasts the plant that carried it, whereas rent-extraction sustains volume, employment, and profit while leaving the underlying capability where it started. In the long run this shows in productivity (Solow, 1957), but the signature is a lagging and imperfect one: productivity can fall while genuine upgrading is under way, as learning curves absorb the gain, and can rise for reasons that have nothing to do with capability, such as capital deepening. Capability accumulates, moreover, through channels that firm accounts do not record.

A set of self-reinforcing mechanisms may hold an economy at the rent-extraction pole. A dependent market economy (Nölke & Vliegenthart, 2009) grows through foreign direct investment into foreign-owned value chains and specialises in low-value assembly. Its foreign owners prefer cheap labour, low R&D, and tax concessions to local upgrading. Antalóczy & Sass (2026) document the resulting value-chain narrowing, in which high-value functions are never relocated and follow-on investment meets diminishing returns. Section 3.3 returns to why these mechanisms are stable.

### 3.2 The four ideal cases

The two axes partition the strategy space into four quadrants, each a limiting case in which the two dimensions are pushed to their extremes. In practice, most strategies lie somewhere in the interior. I name each by its two coordinates (Figure 1).

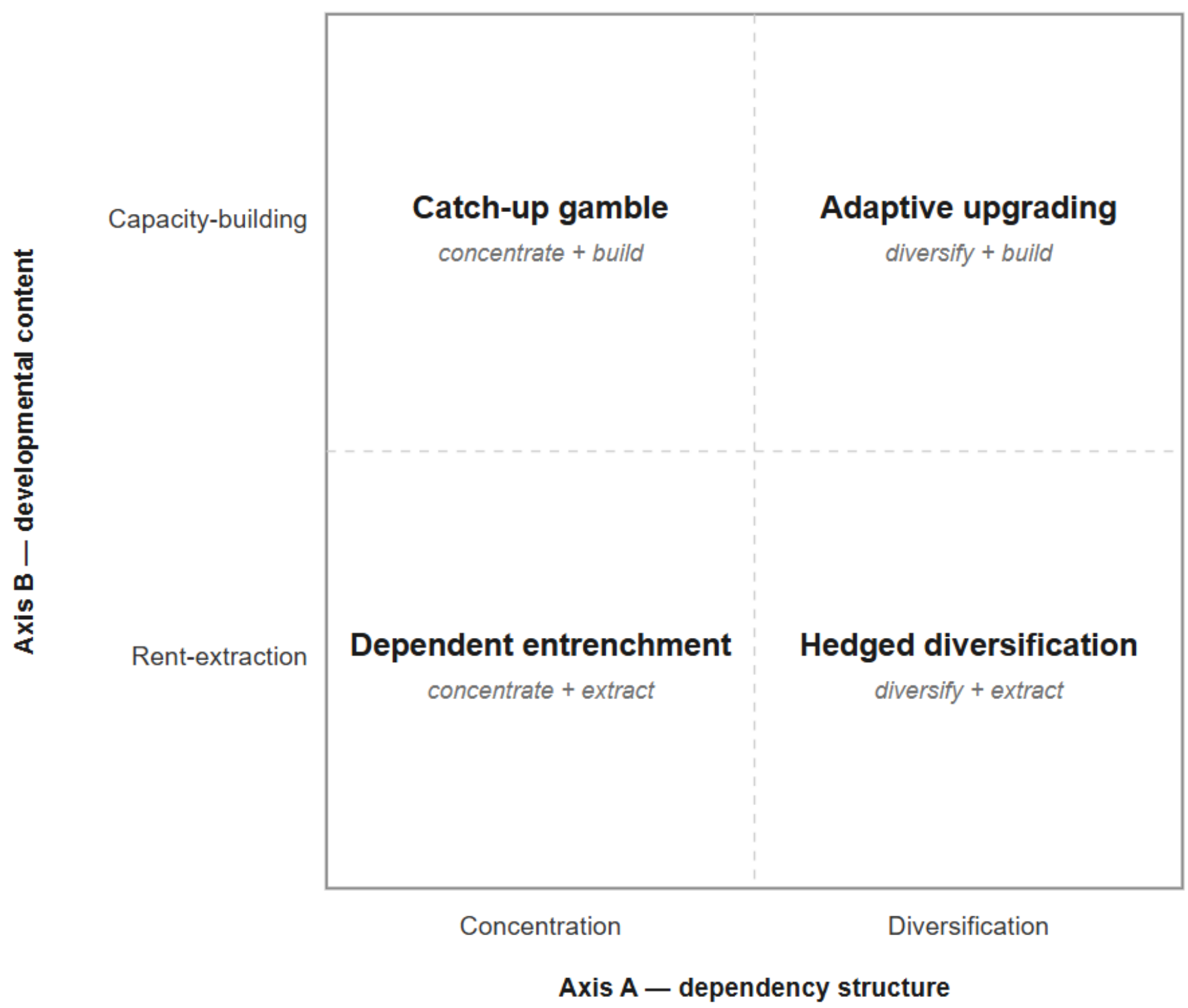

Figure 1. The strategy space defined within the exposure-capability framework, with the four ideal cases at its corners.

Source: own work

Adaptive upgrading (diversify + build) is the unambiguously favourable corner: the economy spreads its dependence across several channels and its growth is built on rising productive capability.

Catch-up gamble (concentrate + build) accepts concentrated dependence as the price of building capability. A country may bet heavily on a single supplier, partner, or technology because that is the fastest route into a new industry, wagering that the capability it gains will eventually let it diversify from a position of strength. The strategy is genuinely developmental but structurally exposed: it pays off only if the capability materialises before the concentrated dependence can be turned against it.

Hedged diversification (diversify + extract) is the mirror image: the economy spreads its dependence and so secures resilience, but its political economy is geared to extracting rents rather than to upgrading. Insured against any single chokepoint, it nonetheless remains developmentally hollow: firms shielded from competition have little reason to innovate, so the diversified base generates volume and profit without the productivity growth that would raise long-run capability.

Dependent entrenchment (concentrate + extract) combines the disadvantages of both: dependence is concentrated, exposing the economy to chokepoint coercion, while its incumbents secure their returns by entrenching against competition rather than by upgrading, so that exposure buys no developmental progress.

The corners are limiting cases, but the axes are continuous. Every economy has some degree of dependency concentration and some developmental orientation, and a single country may sit at different points in different sectors and at different moments. Because positions are points rather than boxes, a strategy can be placed relative to the poles — more or less concentrated, more or less rent-oriented — without claiming an exact coordinate.

### 3.3 Why dependent entrenchment persists

A strategy can occupy any corner of the space, but the corners are not equally stable. Dependent entrenchment is the least desirable corner but it is a self-reinforcing one, too: an economy that arrives there tends to stay, and incremental reform tends to fail. Several literatures that rarely meet — the economics of rent-seeking, the political economy of institutions, development, and the resilience science behind the polycrisis frame — describe this same structure, and their convergence is the strongest reason to treat the corner as a genuine trap rather than a passing condition.

The first source of stability is political: each coordinate of the corner is defended by the actors who profit from it, and they cannot simply be bought off. On Axis A, a hub-holder's leverage depends on its target's continued concentration, so the foreign capital that controls the chokepoints resists diversification. On Axis B, the rent coalition of Scheiring's (2021) ‘accumulative state’ — the political elite, the national champions, and the multinational capital that shares in the arrangement — resists upgrading, because competition would erode the protected positions from which rents flow; and because the same foreign capital often sits in both roles, the two resistances reinforce one another. In the institutional economics literature, Acemoglu (2006) supplies the underlying logic: elites choose and preserve inefficient institutions because the efficient alternative would weaken the power their rents rest on, while commitment problems stop anyone from credibly compensating them to give that power up. The institutions that hold the corner in place are thus not an accident to be corrected but a choice to be defended, which is why escaping it requires a discontinuous push rather than marginal reform.

The second source of stability is economic, a positive feedback internal to rent-seeking. Murphy et al. (1993) show that rent-seeking exhibits increasing returns: as more resources flow into capturing rents, the return to productive and innovative activity falls faster than the return to rent-seeking, so rent-seeking becomes more attractive. The result is a stable bad equilibrium of high rent-seeking and low output. This is the structure Lawrence et al. (2024) call a harmful but stable basin of attraction — a configuration whose damage comes from its durability, held in place by feedbacks that return the system to the same depressed state after any disturbance.

That the trap is more than a theoretical possibility is clear from the historical record: Acemoglu et al. (2001) find that the extractive institutions imposed on many former colonies centuries ago still depress their development today. In its specifically developmental form, the middle-income trap presents a similar condition, where an economy that grew on factor accumulation and cheap labour cannot make the jump to innovation-led growth (Gill & Kharas, 2007).

A country in dependent entrenchment is thus held in place twice over — by the feedbacks that keep it from drifting out, and by the obstacles that meet any deliberate attempt to climb out. The mutual reinforcement described here is a proposition of the framework, resting on the convergence of the literatures above and on institutional evidence, not a relationship this paper estimates; the empirical sections test where a strategy sits.

### 3.4 What the model adds

The two dimensions the exposure-capability model separates have each been studied at length, but in different literatures and rarely together. The external dimension — how concentrated a country's dependencies are, and how exposed that leaves it — belongs to international relations: to the alignment choices of balancing, bandwagoning, and hedging (Kuik, 2021; Walt, 1987) and, more recently, to the analysis of weaponised interdependence (Farrell & Newman, 2019). This work says a great deal about Axis A while remaining largely silent on whether these external dependencies build or erode domestic capability.

The developmental dimension belongs, in the first instance, to economics: to the account of capability accumulation and development traps in the economic-complexity literature (Hidalgo et al., 2007), to the varieties-of-capitalism treatment of dependent market economies (Hall & Soskice, 2001; Nölke & Vliegenthart, 2009), to the institutional-trap literature (Acemoglu, 2006), and to the middle-income trap tradition (Aghion & Bircan, 2017; Gill & Kharas, 2007). This body of work provides a sophisticated view of Axis B but generally treats external dependence as a static backdrop rather than an active strategic variable.

Within the developmental dimension, a second line has asked whether foreign-owned production builds capability in the host economy, and it has located the answer in the channels running between the foreign firm and the domestic economy rather than inside the foreign firm itself: backward linkages to local suppliers, horizontal spillovers to domestic competitors, the mobility of skilled labour, and the absorptive capacity that determines whether domestic firms can use what leaks. The distinction is the one Bell and Pavitt (1993) draw between expanding production capacity and accumulating technological capability, and this literature has consistently found that the first does not deliver the second automatically. Meta-analytic evidence for the transition economies finds forward spillovers statistically insignificant once publication bias and misspecification are accounted for (Fan et al., 2022), and surveys of the region report that indirect effects are not clear-cut and often turn negative in the early years after entry (Hanousek et al., 2011). Firm-panel evidence across eight Central and Eastern European countries finds technology transferred mainly through direct foreign linkages, with horizontal spillovers to domestic firms absent or negative — Hungary among the null cases (Damijan et al., 2003).

This literature also specifies the conditions under which the channels operate, and the specification matters for the case at hand. Spillovers to local suppliers are weaker from wholly owned subsidiaries than from projects with shared domestic ownership, both because local partners lower the cost of finding domestic suppliers and because parents transfer less sophisticated technology to affiliates they do not wholly control (Javorcik & Spatareanu, 2008). Local sourcing rises with the years a subsidiary has operated in the host country and with the autonomy of its local management, and falls where the subsidiary's skill mix reflects firm-specific assets retained at the parent (Amendolagine et al., 2013). For green investment specifically, cross-border acquisitions transfer innovative capability more effectively than newly established greenfield subsidiaries (Amendolagine et al., 2023). And where absorption has been demonstrated in this region, it has been demonstrated in samples of firms far older than the ones examined here (Shkolnykova, 2026). Greenfield entry, wholly foreign ownership, externally directed management, and a few years of operation describe a configuration under which this literature does not expect capability to form. Each of these traditions illuminates one axis while leaving the other in shadow.

The historical exception to this analytical separation is classic dependency theory. Cardoso & Faletto (1979) located associated-dependent development at the intersection of external subordination to foreign capital and the internal alliances that sustained it. Domestic elites, on their account, form structural coalitions with transnational capital to secure their own positions — the rent coalition of Section 3.3, in an earlier idiom. Dependency theory cast a country's position as a structural condition, fixed by its place in the world-system; this model treats it as a space of strategic choice, in which a position can be deliberately moved. It also predates the network mechanisms through which interdependence is now weaponised. This model builds them in: the chokepoints and panopticons of Axis A give external dependence its contemporary coercive edge (Farrell & Newman, 2019).

What the framework adds is the joining of the two axes, and a question that can be answered before the outcomes are in. The spillover literature asks whether capability forms, and answers it well, but only in retrospect and with the long panels that identification requires. The question here is prior to it and differently timed: what position did a strategy purchase, and on what terms? Where in the chain the targeted firms sit, how much of the value they retain, whether that share rises as they mature, and whether the expansion is carried by its own returns — all of this is observable from statutory accounts within a few years of entry, as are the conditions the state attached to its support. A strategy that has bought the lowest-value stage of a chain on terms requiring volume and nothing else has already settled much of what determines whether capability can form later. That is not a measure of capability formation, and I do not claim it as one; it is a measure of the position a state has purchased, available when the commitment is made rather than a generation afterwards.

## 4. Hungary's starting position and the battery bet

I now apply the exposure-capability typology to Hungary. This section locates the economy's starting position on the two axes, sets out the three dependencies that determine it, and discusses the strategic move meant to improve it.

### 4.1 Current position

Hungary is a paradigmatic dependent market economy (Nölke & Vliegenthart, 2009): its growth is financed by foreign direct investment and organised around foreign-owned export sectors. The country's comparative advantage lies in the assembly of semi-standardised goods rather than in their design. The economy is characterised by low domestic R&D, and by a workforce treated as compliant, low-cost labour with little institutionalised voice in firm decisions (Antalóczy & Sass, 2026), unlike the co-determination of coordinated economies such as Germany (Nölke & Vliegenthart, 2009). On the FOI typology of OECD economies, Hungary falls in the dualistic variety, whose development strategy rests on the use of external resources — foreign capital and technology — in an export-led model (Bartha, 2024).

On the developmental axis, the profile of the country sits toward the rent extraction pole. Growth has come from expanding routine production capacity rather than from accumulating the dynamic technological capability that produces upgrading; follow-on FDI meets diminishing returns and high-value functions are not relocated — the value-chain narrowing that Antalóczy and Sass (2026) describe. The competitiveness it sustains is the low-road kind — built on low costs, subsidised FDI, and a weak innovation base rather than on quality and productivity — which is the form of the middle-income trap into which both Győrffy (2022, 2023) and the FOI analysis (Bartha, 2024) place Hungary. The productivity record is consistent with that pole: growth has remained minimal (Muraközy et al., 2018), on the order of one per cent a year over the past decade. Strong investment raises output mainly by adding employment rather than capability (Csaba, 2025). The pattern reaches into the high-technology sectors themselves: even in automotive manufacturing, the specialisation is in low-value-added activity (Győrffy, 2022).

### 4.2 The triple dependency

Hungary's external dependence runs along three axes. The first, and the heaviest, is its integration into German-centred production networks: the automotive and electronics value chains tie Hungarian output to the investment and sourcing decisions of a small number of German original-equipment manufacturers (Győrffy, 2023; Pavlinek et al., 2017). This Western integration remains dominant in aggregate — on Csaba's (2025) account about four-fifths of Hungary's trade and capital flows are still with the Western alliance.

The second axis is energy, where the concentration is on a single external supplier. Despite some diversification of the energy mix, Hungary remains heavily reliant on Russian imports: in 2024 around three-quarters of its gas and oil consumption was imported, with roughly 74 per cent of gas and 86 per cent of oil sourced from Russia — up from 64 and 66 per cent in 2018 (Malak et al., 2025). That dependence is compounded by the structure of delivery, which runs through a few critical transit routes — TurkStream through Serbia and the Adria pipeline through Croatia — while alternative supply from Azerbaijan and Turkey remains limited in scale. The transit structure is itself a chokepoint. The exposure matters all the more because Hungary is among the most energy-intensive economies in Europe, with energy-hungry chemicals, metals, and automotive sectors, so that an EU-wide cut-off of Russian gas could impose output losses exceeding four per cent of GDP (Di Bella et al., 2024).

The third axis, and the most recent, is East Asian battery foreign direct investment: a cluster of Chinese and Korean gigafactory projects. These three exposures all route a critical dependency — production, energy, and now the capital and technology of the EV transition — through a single external partner or bloc. Gagyi and Gerőcs (2025) capture the resulting position when they describe the strategy as turning Hungary into a node that connects German carmakers, East Asian battery producers, and Russian energy — a bridging role the government presents as an opportunity opened by geopolitical realignment.

### 4.3 The strategic move: Eastern Opening, economic neutrality, and the battery bet

The government has responded with an active strategy to improve the country's standing, articulated over the past decade through three connected commitments.

The first is the Eastern Opening. Promoted since the early 2010s and later paired with a Southern Opening, it sought to broaden Hungary's economic and diplomatic ties beyond their Western anchor toward the faster-growing economies of the East — China above all — on the premise that overreliance on Western markets left the country exposed and that new opportunities lay in the emerging world (Csaba, 2025). In the critical political-economy reading, the same turn toward Russia and China served to diversify the sources of development away from Western channels and to support the expansion of domestic capital (Gagyi & Gerőcs, 2025).

The second, and more recent, is the doctrine of economic neutrality. On the government's own account, it meant that Hungary was willing to trade and maintain good financial relations with East and West alike, and to accept investment from any source so long as it is competitive (Csaba, 2025). It was presented as the economic counterpart of a broader emphasis on national sovereignty — that the country's economic decisions should be insulated from outside pressure.

The third commitment, and the centrepiece, is the battery bet. Faced with the electric-vehicle transition, which threatened to strand a car-export model built around the internal-combustion engine, the government moved to make Hungary indispensable to the German auto industry's electric future by sponsoring, at scale, the creation of a domestic battery-manufacturing industry — courting Chinese and Korean cell-makers to build gigafactories alongside the established German assemblers, with the stated ambition of turning Hungary into one of Europe's leading battery producers (Gagyi & Gerőcs, 2025; Ricz & Éltető, 2025). Framed as green reindustrialisation and as a step up into the value chain of the coming automotive economy, the bet is the government's attempt to convert concentrated dependence into new productive capability. In the terms of the typology, the strategy is an attempt to climb out of dependent entrenchment along the developmental axis, accepting and even deepening concentrated dependence in exchange for capability.

**4.4 Catch-up gamble or dependent entrenchment?**

The strategy set out in 4.3 admits two readings, and they correspond to two cells of the typology. On the government's account, the bet is a genuine catch-up gamble: concentrated dependence is accepted as the price of entering the electric-vehicle value chain, and the capability acquired will, in time, justify the exposure and let Hungary diversify from a position of strength. On the critical account, it is dependent entrenchment in a developmental disguise: it deepens concentrated dependence — now on Chinese and Korean battery capital as well as German assemblers and Russian energy — while delivering little of the capability that distinguishes a gamble from a trap.

The critical reading has several strands. The most direct is Csaba's (2025): the government's flagship priority investments — CATL, BYD, BMW, the Záhony transshipment hub — sit at the bottom of the smile curve, in assembly rather than in the design, software, or systems integration where value accrues. The need is not for more such investment but for higher value-added investment and the bet therefore risks reproducing, at larger scale, the low-value-added pattern that already defines the dependent-market-economy model. Csaba (2025) also questions the premise: the decade-long Eastern and Southern Opening to which the bet belongs has, he argues, largely failed. Ricz and Éltető (2025) reach a convergent conclusion from the development-strategy side: the approach does not fit a genuine developmental state, is better understood as a combination of dependency and illiberalism, and amounts to a misapplication of green industrial policy. Gagyi and Gerőcs (2025) read the same move as dependent development under another name — state resources deployed to expand domestic capital within externally set constraints — and warn that flagship projects which consume state support without generating hard-currency returns may strain state capacity rather than build autonomous capability. On the comparative evidence of the FOI typology, the post-2004 EU members sit on the wrong side of a green growth curtain from the coordinated economies of the EU-15 (Bartha, 2024), so the green-reindustrialisation framing is advanced from the structural position the comparative record finds least equipped to deliver it.

These readings are not idle: the critical political-economy literature on Hungary has long described the bet's surrounding architecture — the state–capital coalitions, the rents, the captured procurement — but largely in qualitative, regime-level terms. What it has not done is put the decisive question — whether the bet builds capability or entrenches dependence — to a firm-level empirical test. That test is what separates the two readings: a genuine catch-up gamble should show the targeted firms moving up the chain and retaining a rising share of the value that passes through them; dependent entrenchment should show expansion in volume, employment and profit without that movement.

### 4.5 Hypotheses

The framework makes the disagreement of 4.4 decidable. The bet's exposure-axis position is not in question: it deepens concentrated dependence, and the live alternatives are therefore catch-up gamble and dependent entrenchment. Exposure is a property of the country; what separates the two readings is the position of the firms the strategy targeted, and each implies a different one.

The criteria below follow from the framework. A catch-up gamble is concentrated exposure joined to capacity-building, so if the bet is one, the targeted firms should be moving up: capturing a larger share of the value passing through them, closing on the incumbent base, and shifting toward stages where more of the value sits. Dependent entrenchment is concentrated exposure joined to rent-extraction, so if the bet is that, expansion should proceed without those movements, and the returns should track support rather than production. Each marker is stated so that it could fail: a rising value-added share, a narrowing gap, and a shift up the chain are all things the accounts would record if they occurred.

H1 (Expansion). The targeted firms expand substantially in real output, employment, and fixed investment, at rates well above the comparison groups.

H2 (Catch-up gamble). The bet is building position: the targeted firms are moving toward higher-value stages of the chain and capturing a rising share of the value passing through them.

H3 (Dependent entrenchment). The bet reproduces the existing position at a new product: the targeted firms occupy the same low-value, foreign-directed stage as the incumbent export base, without moving from it.

H4 (Rent mechanism). The expansion is underwritten by state support rather than by the firms' own returns.

Table 1. Hypotheses and the patterns that would support or count against them

| | **Supported if** | **Counted against if** |
|---|---|---|
| **H1** | Real output, employment and assets grow far faster than in the comparison groups | Growth in line with the comparison groups, or confined to one indicator |
| **H2** | Value capture rises over time beyond the recovery from construction losses; the gap to the foreign-owned base narrows from below; activity shifts toward higher-value stages of the chain | Value capture plateaus at or below the base once plants mature; the stage composition of output is unchanged |
| **H3** | Output remains concentrated in the lowest-value stage of the chain; the stage composition does not shift over time | Activity moves toward higher-value stages |
| **H4** | Support intensity far exceeds what comparison firms receive; firms expand while loss-making; the conditions attached require volume rather than capability | Support is comparable to what other firms obtain; firms finance expansion from their own returns; conditions require research, linkages, or technology transfer |

Source: own work

Two features of these criteria deserve note. First, they are directional rather than numerical: no threshold separates a rising value-added share from a flat one, and imposing one would create a precision the data do not carry. Second, H2 and H3 are not simply complements, because both could fail — a bet might expand without moving up the chain while also earning its keep unaided.

## 5. Empirical analysis

This section puts the question of whether Hungary's battery bet is a catch-up gamble or dependent entrenchment to a firm-level panel.

### 5.1 Data

The empirical base is a firm-level panel assembled from the statutory annual financial statements of Hungarian companies, as compiled in the Crefoport database. For each year from 2010 to 2024 I take the ranking of the 1,000 largest firms by sales revenue, and to this comparison universe I add a targeted set of twenty-five firms in the electric-vehicle battery value chain, drawn from Czirfusz's (2023) inventory of the sector and pulled individually irrespective of rank. With the three cell-makers already present in the 1,000 largest firms, the group totals twenty-eight. The combined panel covers 2,240 firms and 15,348 firm-years. It is unbalanced by construction: a core of large firms appears in every year, while many cross the revenue threshold in only some years, so entry and exit reflect movement across the cut-off rather than birth and death.

Each firm-year carries fifteen reported fields: identifiers (registration number, name, year of establishment, four-digit NACE Rev. 2 activity code); the core income-statement lines (sales revenue, export revenue, material-type expenses, personnel costs, depreciation, operating profit, and pre-tax profit); total assets; the average statistical headcount; and two channels of state support, competitive grants and public procurement. The inclusion of the cost lines allows value added to be measured directly. All nominal values are deflated to constant 2024 forint using the implicit GDP deflator published by the Hungarian Central Statistical Office (KSH).

Three limitations should be stated at the outset. Employment is the average annual headcount as reported in each firm's accounts; a small number of records carrying implausible values were treated as missing. The public-procurement field proved unreliable, carrying cumulative and evidently erroneous entries, and is not used; the analysis relies on the competitive-grant field, supplemented for the battery firms as described below. And the battery group, while complete relative to the published inventory of the sector, is small and recent; this constrains the inferential weight any single test can bear and motivates the triangulation strategy throughout.

### 5.2 Sample and group construction

I assign every firm to one of four groups by its four-digit activity code and, for the battery group, by name. The design is deliberately four-way rather than treated-versus-rest, so that the battery firms can be read against both the established foreign-owned export base and the sheltered domestic economy.

The battery-bet group comprises firms in the EV battery value chain, identified by matching company names to Czirfusz's (2023) inventory and supplemented by the cell-makers already present in the revenue ranking (Samsung SDI, SK On, SK Battery). Membership is assigned by name rather than by activity code because the firms are registered across many NACE classes — cell manufacture (2720), separator and foil production (22, 24), components (2453, 2931–2932), and recycling (3821) — so a code-based rule would simultaneously miss battery firms and sweep in unrelated incumbents. German original-equipment assemblers (Audi, BMW, Mercedes-Benz) are excluded from the group and treated as part of the established export base. Within the group each firm is tagged by value-chain stage — cell, material, component, recycling — to permit the position-on-the-chain test. The group contains twenty-eight firms and 394 firm-years.

The Western-dominant group comprises the foreign-investment-led export-manufacturing divisions (NACE 20–22 and 24–29), net of the battery firms. Since the financial statements do not record ownership, I use activity sector as a proxy, following the dependent-market-economy literature that documents the foreign ownership of these sectors (Győrffy, 2022; Nölke & Vliegenthart, 2009; Pavlinek et al., 2017). The Western-dominant group is the most demanding benchmark available, since it is the incumbent export base the bet is meant to extend.

The domestic-insulated group comprises sheltered, largely non-tradable and domestically owned activities — agriculture (01–03), utilities (35–36), construction (41–43), finance (64), and real estate (68) — the part of the economy the dependent-market literature treats as insulated from the export circuit and prone to rent-seeking (Scheiring, 2021; Voszka, 2018; Bartha, 2026). The remaining firms form a broad-market baseline.

The treated and comparison groups are drawn from different selection rules, since the battery sector cannot be recovered from the ranking alone. The variables, however, are identical across both, and Section 5.10 reports a specification imposing a common threshold.

**5.3 Measures and analytical strategy**

The central measurement choice concerns position. Because the panel carries the cost lines of the income statement, I measure value added directly, as the sum of personnel costs, depreciation, and operating profit. From it I form two indicators. Value added as a share of revenue (value added ratio) measures value capture — position on the value chain, and the indicator on which most of the analysis rests. Real value added per worker measures labour productivity in the conventional sense (for the reasons given in 3.1.2, it carries less weight).

Expansion is measured by real sales revenue, employment, and the real total-asset base (the last standing in for investment, since for the gigafactories the asset base is dominated by plant under construction). The rent channel is measured by state support relative to the firm's scale. Here the firm accounts capture only part of the story: the competitive-grant field records ordinary, tender-based support but is blind to the discretionary cash grants awarded by individual government decision (EKD) that finance the large battery investments. For the comparison groups the recorded field is adequate; for the battery firms I supplement it with the documented subsidy record.

Two further ratios characterise the firms' position. Export intensity (export sales as a share of revenue) serves partly to verify the group construction: the Western-dominant firms sell abroad almost entirely, the domestic-insulated firms scarcely at all. Profitability is measured as the pre-tax margin and the return on assets.

The analytical strategy is dictated by the shape of the treated group. The battery bet group is small, young, and dominated by two firms — conditions that dictate a comparative rather than an estimative design. I therefore proceed by triangulation: I read each hypothesis through four complementary lenses and rest the conclusion on their agreement rather than on any one. Section 5.10 supplements these with panel specifications that condition on firm characteristics, and reports the robustness checks.

The first lens is the descriptive trajectory — each real indicator tracked over time as a group median, the battery firms against the comparison groups. It shows the broad direction of travel, but it cannot by itself separate genuine maturation from a compositional effect, the movement in a group's median that merely reflects which firms enter the sample in each year.

The second lens, an event-study on firm age, addresses that limit directly. The battery firms began producing at different dates — staggered across 2019–2024, with four, including the CATL megaproject, still pre-revenue at the close of the panel — so any calendar-year average blends firms at wholly different stages of build-out. I therefore date each firm's entry to its first year of positive revenue and realign the panel on years since entry, comparing every firm at the same age. This isolates real improvement as a plant matures from the mechanical recovery out of the construction-phase losses.

The third lens is cross-sectional benchmarking: battery value capture measured against the foreign-owned export base, and against the other value-chain stages within the battery group itself. Because the battery firms are some two decades younger than the incumbents, comparing them directly with the full base would understate them; where the comparison turns on maturity I benchmark instead against an age-matched slice of that base — its firms founded from 2015 on.

The fourth lens concerns state support: the documented support flowing to the battery firms set against what the comparison groups record, which establishes whether the bet runs on terms any firm could obtain or on terms particular to it. The conditions attached to that support are examined separately, in Section 5.8.

Three points on inference. Where I test a difference statistically, I first collapse each firm to a single value, its median across producing years, so that the test compares firms rather than firm-years and is not inflated by treating one firm's repeated observations as independent evidence. On these firm-level summaries I use the Mann–Whitney rank-sum test, which compares distributions without assuming normality and is unmoved by the outliers that dominate this group, and I report bootstrap confidence intervals for the difference in medians. These tests bear on the level comparisons alone. The trajectory questions — whether value capture is rising or merely holding below the base — are settled by the event-study, not by a significance test. The hypotheses are accordingly stated as claims about position and direction, not as estimated effects.

A word on what this design does not attempt. The obvious alternative would identify the effect of the strategy against a counterfactual. But the counterfactual appropriate to the question is Hungary having made a differently designed bet, since the claims advanced here concern the position the strategy purchased and the terms attached to it. That comparison cannot be constructed from the available variation. Within Hungary there are no untreated firms: essentially every large battery investment received support, so there is no comparison group of unsupported plants against which the supported ones could be set. Across countries, isolating the effect of policy design would require several battery-hosting economies whose instruments differ while their market size, supplier base, entry timing, and position in the European automotive complex are otherwise alike. The region offers too few such cases, and they differ on too many dimensions at once. A country-level synthetic control would not help, since the outcome it could track is a national aggregate, and no aggregate expresses the position of particular firms within a value chain. Section 5.9 makes the comparison that the data do support — Poland's battery chain against Hungary's.

The choice is between a comparative design whose limits can be stated and a causal design whose assumptions could not be defended. Naseemullah (2023) makes the general case that the methods built for micro-level causal inference are ill-suited to questions posed at the scale of national development strategy, and that comparative and historical-structural analysis remains the appropriate register for them. That is the register adopted here. Voszka and colleagues (2023), examining the same support scheme across the whole automotive sector, reach the same practical conclusion, abandoning control-group methods because the largest recipients are systematically unlike the rest of the population. What follows establishes where the targeted firms sit, how they compare with the relevant benchmarks, and what the state required in return. It does not establish what a differently designed strategy would have produced.

**5.4 Descriptive statistics**

The firm distributions are heavily right-skewed, and the battery group in particular is dominated by a few large producers, therefore I report medians rather than means. Table 2 profiles the groups. A firm-year is producing when it reports positive sales revenue and carries the cost lines from which value added is calculated. The distinction matters most for the battery group: of its 394 firm-years only 114 are producing, the rest being the empty pre-production years of firms that have not yet reached revenue. The comparison groups are producing almost by definition. Two features stand out. The battery firms are young — a median year of establishment of 2019, against 1998 for the Western-dominant export base — so they are measured against incumbents two decades more mature, a gap I return to in Section 5.6. And the median producing battery firm is small in revenue (13.5 bn HUF) yet asset-heavy (26.7 bn HUF), an asset-to-revenue ratio near two against roughly 0.7 for the export base. This is the capital intensity of gigafactory production, in which plant and equipment dwarf current turnover.

Table 2. The four groups of the sample: profile (size on producing firm-years, real 2024 HUF)

| **Group** | **Firms** | **Firm-years** | **Producing** | **Median founded** | **Median revenue (bn)** | **Median employment** | **Median assets (bn)** |
|---|---|---|---|---|---|---|---|
| Battery bet | 28 | 394 | 114 | 2019 | 13.5 | 64 | 26.7 |
| Western-dominant | 374 | 3,270 | 3,270 | 1998 | 49.5 | 562 | 34.5 |
| Domestic-insulated | 419 | 2,334 | 2,333 | 2002 | 45.9 | 187 | 58.8 |
| Broad market | 1,419 | 9,350 | 9,347 | 2000 | 39.2 | 192 | 20.0 |

Source: own calculations

Table 3 reports the analytical variables across the groups, on producing firm-years. Battery value added as a share of revenue is lower than the Western-dominant base — a median of 0.14 against 0.18 — and far more dispersed, with a negative lower quartile that reflects the ramp years in which operating losses exceed personnel costs and depreciation combined. The same ramp shows in profitability: the median battery firm-year is loss-making, both at the pre-tax margin (−1.8%) and on assets (−1.0%), where the comparison groups earn a steady three to six per cent. The export-intensity column does double duty, confirming that the group construction captures what it claims: the Western-dominant firms export almost all of their output, the domestic-insulated firms almost none. The battery column is bimodal — the cell-makers export to their parent networks while the supplier firms sell into the domestic gigafactory cluster.

Table 3. Summary statistics by group (producing firm-years; median [IQR])

| | **Value added ratio** | **Profit margin** | **ROA** | **Material ratio** | **Export intensity** |
|---|---|---|---|---|---|
| Battery bet | 0.138 [−0.32, 0.21] | −1.8% [−107%, 4%] | −1.0% [−8%, 2%] | 0.86 [0.78, 1.26] | 0.02 [0.00, 0.99] |
| Western-dominant | 0.178 [0.11, 0.26] | 3.5% [1%, 7%] | 5.6% [1%, 11%] | 0.80 [0.70, 0.87] | 0.92 [0.73, 0.98] |
| Domestic-insulated | 0.146 [0.05, 0.28] | 4.8% [1%, 13%] | 4.1% [1%, 10%] | 0.80 [0.46, 0.91] | 0.00 [0.00, 0.03] |
| Broad market | 0.112 [0.05, 0.23] | 2.6% [1%, 6%] | 5.7% [1%, 12%] | 0.85 [0.68, 0.94] | 0.10 [0.00, 0.58] |

*Note: IQR stands for interquartile range — the spread between the 25th and 75th percentiles*

Source: own calculations

Table 4 sets out the battery group in detail. The group is staggered in entry: first revenue clusters in 2019–2024, and four firms — including the CATL megaproject — were still pre-revenue at the end of the panel. It spans the value chain, but it is overwhelmingly concentrated in cell fabrication by weight: the two largest cell-makers alone accounting for 77 to 84 per cent of the group's real revenue in recent years.

Table 4. Battery group: coverage, composition, and concentration

| **Coverage (firm-years)** | | |
|---|---|---|
| Total firm-years | 394 | |
| With financial data | 242 | |
| Producing (positive revenue) | 114 | |
| **Composition by value-chain stage** | firms | producing firms |
| Cell | 6 | 5 |
| Material | 11 | 9 |
| Component | 8 | 8 |
| Recycling | 2 | 2 |
| OEM (BYD) | 1 | 0 |
| **Entry and concentration** | | |
| First-revenue years | 2019–2024 (clustered) | |
| Firms still pre-revenue at 2024 | 4 (incl. CATL) | |
| Top-2 firms' share of real revenue | 77–84% | |

Source: own calculations

Figure 2 traces the scale of the bet over time. The single firm present before 2017 is Samsung SDI's earlier display-panel operation in Göd, not battery production. The bet emerges from 2018, when the group expands steeply in real output, employment, and assets as the gigafactories are built and come on stream, before real revenue softens in 2024 amid the slowdown in European EV demand.

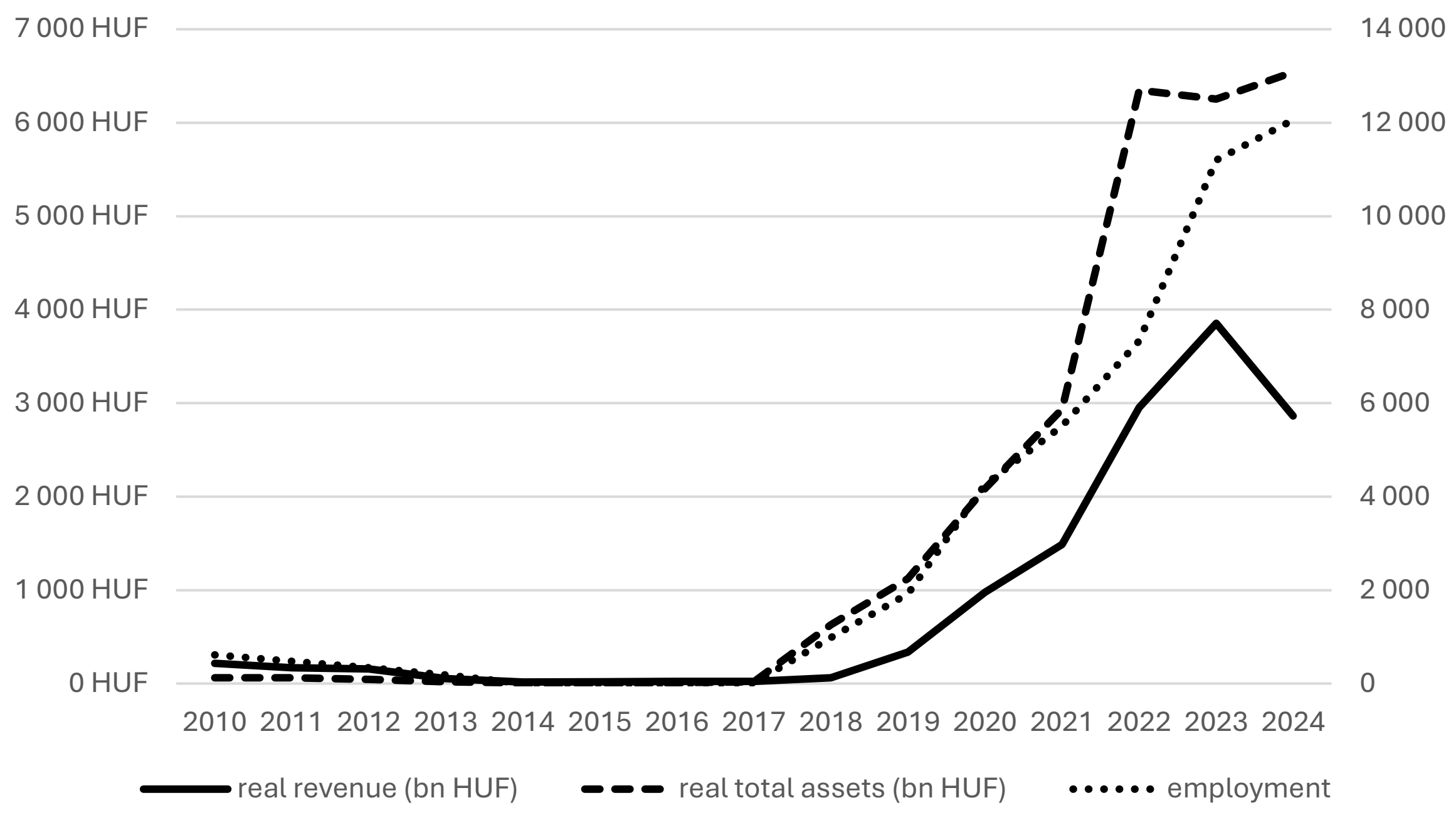

Figure 2. The expansion trajectory: battery aggregate real revenue, real assets (left axis), and employment (right axis), 2010–2024

Source: own calculations

**5.5 Expansion (H1)**

The first hypothesis is the precondition the two readings share: that the bet materialised as large-scale industrial growth. Its marker is straightforward — real output, employment and assets growing far faster than in the comparison groups, and not merely in one indicator. Table 5 sets the battery group's real expansion over the bet period against the three comparison groups.

Table 5. Real expansion by group, over the bet period (constant 2024 HUF, bn; employment in persons)

| Group | Real revenue 2016 → 2024 | × | Employment 2018 → 2024 | × | Real assets 2016 → 2024 | × |
|---|---|---|---|---|---|---|
| Battery bet | 24 → 2,861 | ×119 | 983 → 12,058 | ×12 | 9 → 6,541 | ×692 |
| Western-dominant | 33,770 → 28,551 | ×0.8 | 241,212 → 202,125 | ×0.8 | 27,824 → 26,248 | ×0.9 |
| Domestic-insulated | 14,912 → 28,874 | ×1.9 | 88,260 → 106,895 | ×1.2 | 78,344 → 102,104 | ×1.3 |
| Broad market | 69,318 → 54,311 | ×0.8 | 390,290 → 406,806 | ×1.0 | 70,421 → 45,389 | ×0.6 |

Source: own calculations

The growth is of a different order from anything else in the panel. Between 2016 and 2024 the battery group's real output rose roughly 119-fold, its real asset base 692-fold, and its workforce grew twelvefold from 2018, the first year its employment can be measured. Against this the comparison groups are flat or shrinking in real terms. The contrast with the Western-dominant group is the most telling: the foreign-owned export base the bet is meant to extend did not grow at all over the period but contracted, its real revenue and employment both falling by around a fifth — a reminder that the established model was already losing momentum when the battery investments arrived. The expansion is so far above the comparison that the orders of magnitude speak for themselves.

The scale is substantial in absolute terms too. The battery cluster rose from a negligible share of the sampled corporate universe to 2.5 per cent of its real revenue in eight years, an unusually rapid structural addition concentrated in a handful of plants: Samsung SDI alone went from 34 to over 1,800 billion forint in real revenue.

Real output peaks in 2023 and falls back in 2024 even as employment and assets keep rising; Section 5.10 reports the comparison on a pre-downturn endpoint, where the multiple is larger still. H1 is supported on all three indicators. It is also the least discriminating of the four, since both readings predict it: what separates them is not whether the bet produced growth but what that growth purchased.

**5.6 Value capture: catch-up or entrenchment? (H2 and H3)**

H2 requires three things: value capture rising beyond the recovery from construction losses, the gap to the foreign-owned base narrowing from below, and activity shifting toward higher-value stages. H3 requires output concentrated in the lowest-value stage and a stage composition that does not shift. I present the value-added series three ways — its level against the base, its structure across the chain, and its trajectory over time — because each bears on a different marker, and only their convergence is decisive.

The level comparison is reported in Table 6. On firm-collapsed medians, battery value capture sits well below the full export base — a value-added share of 0.075 against 0.179, a difference comfortable past conventional significance. But the comparison that matters is with the age-matched base, the export firms established from 2015, and there the gap narrows to a third of its size and ceases to be significant: 0.075 against 0.106 ($p = 0.26$). Most of the apparent shortfall is youth, not battery-specificity — the young foreign firms capture little more. Labour productivity tells the same story more faintly, below the base but not decisively. Taken alone, the level evidence is therefore equivocal. It bears on the second marker of H2 — whether the gap to the base is narrowing — and shows a gap to the full base but none to the age-matched one, which is consistent with youth rather than with either reading.

Table 6. Value capture: level comparison (firm-collapsed median over producing years; Mann–Whitney p)

| Indicator | Battery (n=24) | Young Western, ≥2015 (n=27) | Full Western (n=374) | p vs young | p vs full |
|---|---|---|---|---|---|
| VA / revenue | 0.075 | 0.106 | 0.179 | 0.26 | <0.001 |
| Real VA / worker (M HUF) | 14.2 | 23.1 | 17.5 | 0.09 | 0.06 |

Source: own calculations

One caveat attaches to the young Western benchmark. It is drawn from the top-1000 revenue ranking, so it contains only young firms that reached Hungary's largest corporations within a few years of founding, whereas the battery group was pulled irrespective of rank and includes small suppliers. The benchmark is therefore a demanding one, and the comparison is conservative.

The structure of value capture is where the evidence turns sharp. Table 7 breaks the group down by stage and sets each stage's value capture beside its weight in the bet. Eighty-eight per cent of the battery group's real revenue is generated in cell fabrication, and cell manufacture is the lowest-value stage in the group, with a median value-added share of 0.048 — a fraction of what the export base captures. The stages that do capture more, components and recycling at around 0.17, are on a par with the base but marginal, together less than eight per cent of the bet's revenue; the materials firms remain in ramp losses. Value is captured where the chain is climbed, and almost none of the bet is there.

Table 7. The smile curve: value capture by stage

| Stage | Firms | Median VA / revenue | Share of battery real revenue |
|---|---|---|---|
| Cell | 5 | 0.048 | 88.1% |
| Material | 9 | −0.83 | 4.2% |
| Component | 8 | 0.172 | 7.2% |
| Recycling | 2 | 0.175 | 0.5% |

Source: own calculations

The trajectory, finally, tests whether this is a starting point or a destination. Figure 3 aligns the firms on years since first production and traces median value capture from there. Value capture is negative in the entry year, when fixed costs are carried against minimal output, rises through the following two years, and settles from t+2 at between 0.15 and 0.18 — at the level of the export base — where it remains for as long as the panel observes it. What the figure does not show is any continued ascent once that level is reached. The one plant in the region with a longer record, LG Energy Solution's Wrocław facility, is plotted alongside and follows the same course two years further out; Section 5.9 takes up that comparison. The value-weighted series tells the complementary half of the story: weighted by output, the battery group captures around 0.19, comparable to the base. That figure is carried by the two largest producers, which account for the great majority of battery revenue, so it describes those plants rather than the sector; Section 5.10 reports the check.

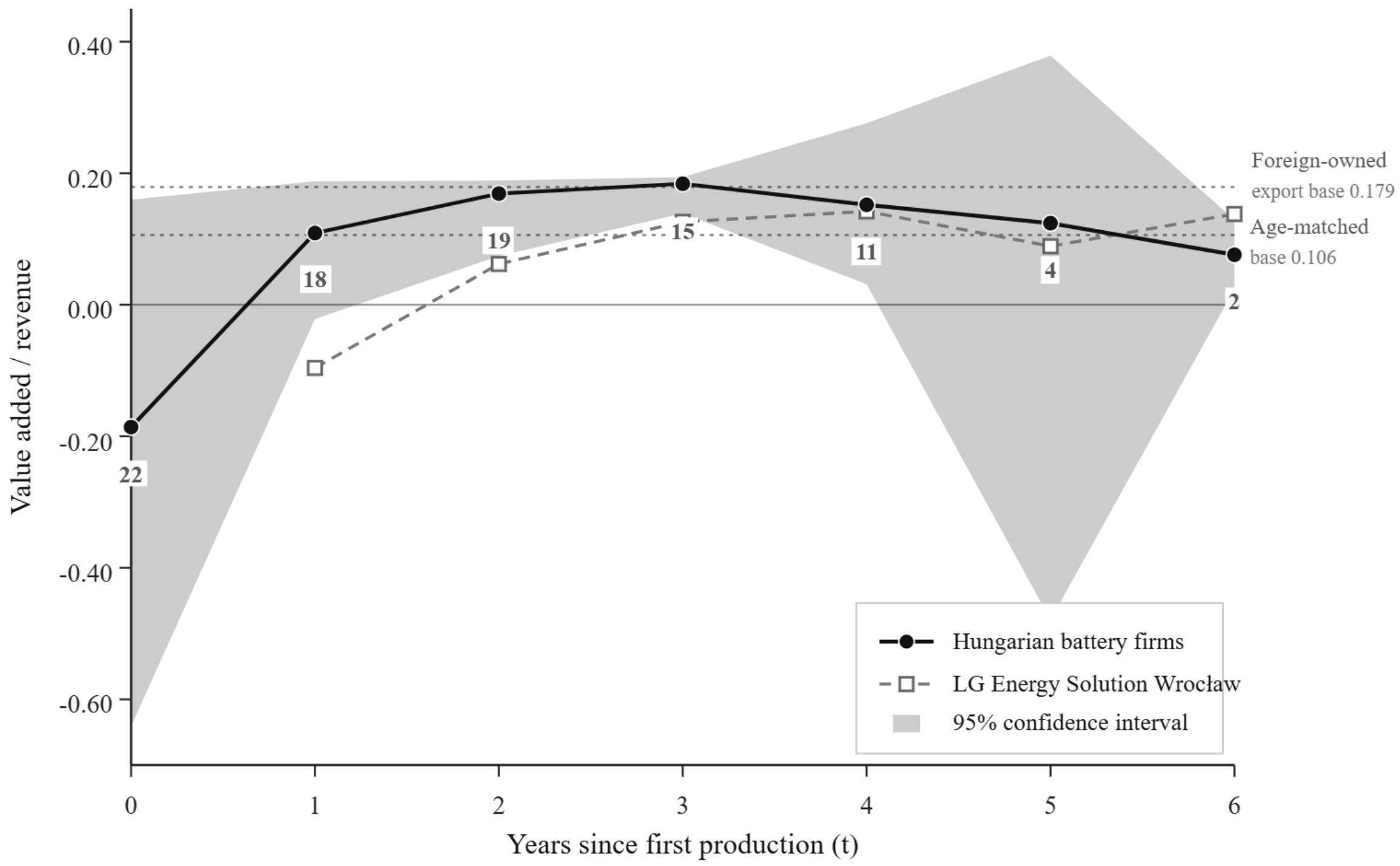

Figure 3. Value capture over the plant life-cycle: value added as a share of revenue, by years since first production

*Notes: Hungarian battery firms show median across firms at each event year, with bootstrap 95% confidence interval. Number of firms at each event year: 22, 18, 19, 15, 11, 4, 2. A firm enters at its first year with revenue above 100 million forints. LG Energy Solution Wrocław is a single firm, shown from its first full production year (2019); see Section 5.9. Reference lines give the firm-collapsed medians for the foreign-owned export base and its age-matched subset.*

Source: own calculations

The three readings can now be set against Table 1. H2 fails on all three markers. Value capture does not rise beyond the recovery from construction losses: the event-study climbs out of the ramp and then plateaus. The gap does not narrow progressively: value capture reaches the age-matched base and remains there. And the stage composition does not shift, cell fabrication remaining some seven-eighths of output throughout.

H3 holds on each marker. Output remains concentrated in cell fabrication throughout and the stage composition does not shift. One result deserves note because it bears on how H3 should be read: restricted to plants with three or more producing years, battery value capture is statistically indistinguishable from the age-matched base (Section 5.10). The firms are not capturing too little value for what they are, but what they are is the lowest-value stage of the chain, and they have not moved from it over the period observed. The bet, in short, neither raises nor lowers the position of the model it joins; it substitutes within it — battery fabrication for automotive assembly, at the same level of value capture.

**5.7 The rent mechanism (H4)**

If the battery firms capture no more value than the export base and remain at the bottom of the chain, what sustains an expansion of the magnitude H1 recorded, through years of losses? The fourth hypothesis answers that the bet is held in place by state support of a kind and scale unavailable to other firms. The evidence has three parts: that the support is disproportionate, that the firms do not stand on their own without it, and — taken up in Section 5.8 — what the support required in return.

Table 8 shows where ordinary state support goes. Through the recorded channel — competitive, mostly EU-cofinanced grants — the Hungarian state subsidises broadly: the sheltered domestic economy draws nearly half of all recorded grants, the broad market the other half, and the foreign export base draws grants in one year out of ten, if in smaller amounts. This is the dual subsidy regime the dependent-market literature describes: the state underwriting both its domestic clientele and the FDI sector that anchors the export model (Bohle & Greskovits, 2012; Győrffy, 2023; Scheiring, 2021). The battery firms, by contrast, draw essentially nothing through this channel.

Their support runs through a different channel. The large battery investments are financed by discretionary cash grants awarded by individual government decision (the EKD scheme), supplemented by tax allowances, land, and dedicated infrastructure — instruments that do not appear in the firms' competitive-grant line. The documented record of these awards is of an altogether different order from the grants of Table 8: Samsung SDI received an EU-approved €89.6 million for its Göd expansion; SK On €209 million for its Iváncsa plant; CATL a reported €800 million for Debrecen; Toray €46.5 million for separator film (European Commission, 2023, 2024a; Tagliapietra et al., 2025; Transport & Environment, 2025). Documented cash grants to the battery firms amount to some €1.6 billion, around 640 billion forints: roughly a fifth of all the recorded grants paid to the two-thousand-odd comparison firms over fifteen years. Counting tax allowances, land, and dedicated infrastructure alongside the grants, total public support committed to the sector by 2025 approaches 1,500 billion forint, close to two per cent of GDP. Measured against investment the awards run at seven to fifteen per cent, against the half a per cent of revenue that ordinary recipients receive; measured against jobs, the support approaches 124 million forint, some €311,000, for each position created. For comparison, Bartik (2020) puts the cost of customised training and technology-consulting programmes at around $34,000 per job created, and of tax-incentive inducement — the least efficient of the common instruments — at roughly $196,000.

Table 8. State support: what each group records, and what the bet actually received

| Group | % firm-years with a recorded grant | Total recorded grants (bn HUF) | Median grant / revenue |
|---|---|---|---|
| Battery bet | 0.5% | ≈ 0 | — |
| Western-dominant | 10.6% | 143 | 0.52% |
| Domestic-insulated | 8.2% | 1,366 | 0.58% |
| Broad market | 5.6% | 1,529 | 0.33% |

Source: own calculations

Table 9 asks whether the firms need it — whether their returns could carry the expansion on their own. They could not. The median battery firm-year is loss-making, both on revenue (a pre-tax margin of −1.8 per cent) and on assets (a return of −1.0 per cent), where the comparison groups earn a steady three to six per cent. The battery firms run pre-tax losses in fifty-seven per cent of their producing years, against fewer than one in five for every other group. A sector expanding more than a hundredfold while losing money at the median is not financing that expansion from its own returns — and this is not merely the construction ramp working itself out. Aligned on firm age, median margin climbs from negative at entry to a bare +1.7 per cent by the third producing year and then slips back below zero; the two mature flagship cell-makers, four to seven years into production, hold between −2 and +2 per cent and both fall into loss again in 2024. The firms do not become profitable as they mature. Their continued operation is financed from outside: by parent-company capital and by the public subsidy.

Table 9. Returns: does the bet stand on its own? (producing firm-years)

| Group | Median margin | Median ROA | Pre-tax-loss years | Negative-VA years |
|---|---|---|---|---|
| Battery bet | −1.8% | −1.0% | 57% | 33% |
| Western-dominant | 3.5% | 5.6% | 18% | 2% |
| Domestic-insulated | 4.8% | 4.1% | 15% | 6% |
| Broad market | 2.6% | 5.7% | 14% | 3% |

Source: own calculations

One qualification tempers the profitability evidence. The losses of 2024 coincide with a sharp downturn in the sector's market: European EV demand cooled, global cell overcapacity pushed prices down, and Hungarian battery output fell by roughly half over the year (Eszterhai & Vörös, 2025). Some of the recent weakness is therefore cyclical, and a fuller verdict on whether the mature plants can earn their keep must await the market's recovery. But the cycle does not explain the structure away. The losses are not confined to 2024, they run through the ramp and the thin-margin years before it.

H4 has three markers, and two of them are settled here. Support intensity far exceeds what comparison firms receive: the documented awards run at seven to fifteen per cent of investment against half a per cent of revenue for ordinary recipients, and reach the battery firms through a channel the comparison groups do not use. And the firms expand while loss-making, which Table 9 establishes at the median and which holds under conditioning — allowing for age, size, sector and year, the battery firms run a pre-tax margin some seven percentage points below otherwise comparable firms, a gap that does not disappear when the downturn years are excluded (Section 5.10). Expansion on this scale is not being financed from the returns the firms generate. The third marker concerns what the support required in return, which is the subject of the next section.

### 5.8 What the support required

Large state support is not in itself evidence of rent. The East Asian catch-up episodes that anchor the developmental-state literature were heavily subsidised, and subsidy was the instrument through which capability was built (Amsden, 2001/2003; Rodrik et al., 1995). What distinguished them from unproductive transfers is the obligations attached to them: Amsden's reciprocal control mechanisms, under which firms received support in exchange for measurable performance standards and lost it when they failed to meet them.

Support under the EKD scheme is governed by Government Decree 210/2014 (VIII. 27.), whose §14(1) sets out the binding conditions: a minimum volume of eligible investment; growth in the wage bill, in net revenue, or in the two combined; maintenance of the investment in the region for at least five business years; at least a quarter of the cost from the firm's own resources; and maintenance of the existing headcount through the operating period. Two of these are enforceable by withdrawal and repayment under §26(2): the headcount obligation and the revenue or wage-bill obligation (Government of Hungary, 2014). None of the binding conditions requires research and development, technology transfer, or the development of domestic suppliers. One absence needs qualifying: the decree could not have required local sourcing even had the government wished it, since EU state aid rules prohibit aid conditioned on the use of domestic goods and services (Article 1(5)(b), European Commission, 2014), and §8(1)(p)–(q) reproduces that prohibition. Capability-building conditionality had to travel through other channels (research obligations, training requirements, supplier-development commitments) and the regional investment track imposes none of them.

The decree is not silent on capability, though. §10 lists considerations for the awarding authority, among them the investor's cooperation in developing a competitive Hungarian supplier base, the share of employees with higher education, and, for research projects, existing Hungarian R&D activity and links to domestic universities. None of these appears among the conditions of §14, and none among the withdrawal triggers of §26. Capability is what the decision-maker may weigh; capacity is what the recipient must deliver on pain of repayment. The design is not simply an oversight. Evans (1995) distinguishes states that are embedded in their industrial base while retaining the autonomy to impose conditions on it from those whose embeddedness is unaccompanied by such autonomy. In the latter, exemption from conditions is part of what the relationship supplies. A scheme that weighs capability at the point of award but never requires it is what that second configuration produces.

The supplier-development criterion is worth a further word, because the ownership structure of these investments bears on what it could have achieved even had it been binding. The battery plants are wholly owned subsidiaries of their foreign parents, with no Hungarian equity participation — the configuration under which local linkages are weakest (Section 3.4, Javorcik & Spatareanu, 2008). A criterion the awarding authority may weigh but need not enforce, applied to firms whose ownership already works against linkage formation, is unlikely to produce much.

The scheme also contains a capability-contingent instrument that the battery investments did not use. §2(1)(b) provides separate support for research and development projects, conditioned under §9(1a) on the creation of at least ten new research positions, half of them requiring higher education, with disbursement tied to meeting those targets. The Korean battery investments were authorised as regional investment aid, not under this track (European Commission, 2023, 2024a). The state possessed an instrument that conditions support on research capacity and did not apply it to this large-scale industrial commitment.

Since October 2019, the creation of jobs (as distinct from the maintenance of existing ones) has not been among the required conditions; the criteria are the wage-bill and revenue thresholds set out above. The largest single award identified in the State Audit Office's review illustrates what this permits: SK On's Iváncsa plant received 76.4 billion forints with no additional-employment commitment attached (State Audit Office of Hungary, 2025). Voszka et al. (2023), examining the same scheme across the whole automotive sector, single out that award on the same grounds. The Audit Office also found that the effects of support on value added were never quantified during the preparation of award decisions. The scheme purchases capacity; it does not ask for capability.

### 5.9 The pattern outside Hungary

Two objections to the foregoing cannot be answered with Hungarian data. The first is that the battery firms are too young: if upgrading takes decades, five years of observation shows little. The second is that whatever the panel finds may be a product of Hungarian policy rather than of the position the firms occupy. Both can be addressed, if imperfectly, by examining the same industry in a country whose support regime differs in construction, but whose structural position does not. Poland hosts the region's largest concentration of battery-chain production, and its plants are on average older than Hungary's. Its support for new manufacturing investment has run principally through corporate income tax exemption: first through permits to operate in the special economic zones, and, since the Act of 10 May 2018, through the decision on support, which extended the instrument nationwide. The contrast with the Hungarian awards documented above is one of form. Poland's principal instrument forgoes revenue, and conditions eligibility on a set of qualitative criteria that enter the decision on support as obligations over the maintenance period rather than as considerations at the point of award (Council of Ministers of Poland, 2018, §5 and Annex 1, Table 2). The difference should not be overstated. Both instruments are regional investment aid, both are driven principally by capital expenditure, and Polish firms select their qualitative criteria from a menu that includes social and locational commitments as readily as research or skills ones. But the two regimes are not the same instrument, and that is sufficient for the comparison that follows.

I identified ten Polish firms in the battery value chain from a published sector inventory (Lewicka et al., 2025) and retrieved their statutory annual financial statements as filed with the Polish National Court Register, in structured form via aleo.com. Polish accounting permits two income-statement formats, and only the nature-of-expense variant separates material costs, personnel costs, and depreciation; where firms file that variant, value added is computed exactly as for the Hungarian panel. Six firms yield usable post-ramp observations (Table 10).

Table 10. Polish battery-chain firms: value capture by stage

| Firm | Stage | First production | VA_ratio 2022 | 2023 | 2024 | Operating margin 2024 |
|---|---|---|---|---|---|---|
| LG Energy Solution Wrocław | cell | 2018 | 0.142 | 0.089 | 0.138 | −1.5% |
| BMZ Poland | component | 2010 | 0.180 | 0.170 | 0.185 | 6.2% |
| Northvolt Poland | component | 2019 | 0.151 | 0.114 | — | — |
| SK hi-tech Battery Materials | material | 2021 | 0.451 | 0.503 | 0.801 | 8.3% |
| Guotai-Huarong (Poland) | material | 2018 | −0.287 | −0.152 | 0.083 | −11.7% |
| Capchem Poland | material | 2024 | — | — | — | 4.0% |

*Note: Northvolt Poland's 2024 accounts record a write-off of 542 million zloty following its parent's insolvency and are excluded. Capchem's 2024 statement uses the cost-of-sales format, from which value added cannot be derived.*

Source: own calculations

The horizon objection first. LG Energy Solution's Wrocław plant, Europe's largest cell facility, has produced since 2018 — longer than any Hungarian cell plant. Its value capture climbs out of the construction trough from −0.096 in 2019, its first full year of production, to 0.142 in 2022, and goes no higher. The shape is that of Figure 3, the Hungarian event-study, extended by two years. Six years is not the decades that upgrading histories often require, and this cannot be decisive. But it is the longest observation the region affords, and through it the ascent does not resume.

The objection has a further form worth addressing. Capability is generally held to accumulate in stages, and the sequence is slow: in the early phases it works through factor endowments and linkages between sectors, and only later does technological knowledge become a source of advantage in its own right (Bell & Pavitt, 1992; Lall, 1992). The same sequencing organises the World Bank's recent account of escape from the middle-income trap, in which investment precedes the infusion of foreign technology into the wider economy, and infusion precedes innovation (World Bank, 2024). On these accounts, five or six years is not a period in which one would expect to observe accumulated capability, and its absence here would prove little. But the argument cuts both ways. The early phase of that sequence is not innovation but the diffusion of technology into the wider economy, and while the linkages through which diffusion runs lie outside what firm accounts record, its precondition does not. Diffusion presupposes plants that are moving — taking on higher-value stages, retaining more of what passes through them — and neither is happening. The plants sit at the stage they entered. That is not evidence that capability has failed to accumulate, which the window is too short to show. It is evidence that the sequence which would produce it has not visibly begun.

The second objection is more consequential, and the Polish evidence shapes what the paper claims. Value capture in Poland orders itself by value-chain stage much as it does in Hungary: the cell plant lowest, components above it, and the established materials producers higher still. Two support regimes produce the same ranking. Low value capture at the cell stage is therefore better understood as a property of that position in the chain than as a consequence of Hungarian policy in particular. This separates two claims that the Hungarian evidence alone had left joined: the position accounts for the level of value capture, while what Hungarian policy accounts for is the absence of any instrument directed at moving off that position or building capability within it, which is the subject of Section 5.8.

The stage ordering should not be pressed too hard, because variation within a stage is wide. SK hi-tech's value capture rises from 0.45 to 0.80 across four years, while Guotai-Huarong, in production since 2018, remains below 0.10 throughout and turns negative in 2022 and 2023, when material costs exceeded revenue. Stage correlates with value capture; it does not determine it.

The Polish record also confirms that the weakness of 2024 was sectoral rather than Hungarian. LG's revenue fell by 45 per cent that year and the plant moved into loss, as Section 5.7 found for the Hungarian producers. And it shows that investment of this kind carries real risk under a different support regime: Johnson Matthey built a cathode-materials plant at Konin to a fixed-asset value of 1.33 billion zloty, never produced commercially, wrote off almost the whole of it and entered liquidation in March 2022, while SK Nexilis remains pre-production after four years and 2.67 billion zloty of committed assets.

Two limits should be stated. Poland is compared here only within the battery chain: there are no Polish comparison groups equivalent to the Hungarian export base and sheltered economy, so this replicates the within-industry pattern across countries rather than the full research design. And six firms with short series cannot establish a general regularity. What the evidence supports is narrower and sufficient: the value-capture profile documented in Hungary is not peculiar to Hungary, and the stage of the chain a plant occupies predicts that profile better than the country hosting it.

**5.10 Robustness**

The comparisons above invite doubt about whether the value-capture results are artefacts: the battery group is small and dominated by two producers, it enters the sample under a different rule from the comparison groups, its firms are young, and the final year of the panel coincides with a sectoral downturn. I test these, and two questions about measurement, in turn. Table 11 reports the level comparison under four alternative specifications.

Table 11. Value capture under alternative specifications (firm-collapsed medians, producing firm-years)

| Specification | Battery firms | Median VA ratio | p vs full base | p vs age-matched base |
|---|---|---|---|---|
| Baseline | 24 | 0.075 | <0.001 | 0.26 |
| Excluding the two largest producers | 22 | 0.067 | <0.001 | 0.16 |
| Common selection threshold | 20 | 0.080 | <0.001 | 0.33 |
| Mature producers only (≥3 producing years) | 20 | 0.131 | 0.019 | 0.91 |

*Note: The common-threshold specification retains only battery firms large enough to have entered the comparison universe on their own.*

Source: own calculations

The result does not depend on the two flagship producers: dropping Samsung SDI and SK On leaves the median almost unchanged and the comparison with the full base slightly stronger. Nor is it an artefact of the different sampling rules, since restricting the battery group to firms that would have qualified for the comparison universe on size alone gives the same pattern. The fourth specification is the most informative. Restricting attention to firms with at least three producing years raises the median to 0.131 and leaves it statistically indistinguishable from the age-matched base, while still below the full base. That is the paper's claim about maturation, stated in Section 5.6, in the form of a test rather than an inference.

The endpoint matters less than might be expected. Measured to 2023, before the downturn, the battery group's real revenue expansion is 161-fold from 2016 rather than 119-fold, against one to two for every comparison group: H1 holds on either endpoint. The 2024 collapse in battery margins, from a median of +1.4 per cent to −12.7 per cent, has no counterpart in the Western base, whose median margin is unmoved at 2.6 per cent, which is consistent with a demand shock specific to the sector rather than a divergence in underlying performance.

Three further checks bear on measurement rather than sample composition. Because value added is derived from reported accounts, it is exposed to transfer pricing within multinational groups. Personnel costs are far harder to relocate across borders than profit, so the share of revenue paid in wages offers a partial check: restricted to post-ramp firm-years, the battery median is 0.106 against 0.108 for the full base and 0.071 for the age-matched base, neither difference approaching significance. A null result here is what one would want: had the battery firms shown an unusually low wage share alongside their low value added, profit shifting would be the obvious suspect. Separately, a small number of firm-years across the panel report positive employment with no personnel costs, an evident reporting gap; treating these as missing changes no result in Table 11 by more than 0.002.

A second check compares the accounts-based measure with official statistics. Eurostat's structural business statistics report value added as a share of output for the manufacture of batteries and accumulators; for Hungary the series runs at 18.0, 16.2, 12.9 and 21.2 per cent across 2021 to 2024 (Eurostat [sbs_ovw_act]), against the value-weighted ratio of roughly 0.19 computed here. The activity class is narrower than the battery group used in this paper, and the official series begins only in 2021, so the comparison validates the measure rather than replicating the sample. Within those limits, an independently constructed official series and the panel-based measure agree closely, which gives no indication that the accounts-based construction is distorted.

One finding runs the other way. The value-weighted ratio reported in Section 5.6, which shows the sector as a whole capturing roughly the base level of value, is carried entirely by the two flagship producers. Excluding them, the remaining battery firms capture some 1.3 per cent of revenue as value added, reflecting the ramp-stage losses of the materials firms. This is arithmetically unsurprising, since the two account for the large majority of battery revenue, but it means the value-weighted figure describes those two plants rather than the sector. The firm-collapsed median, on which the analysis principally rests, is robust to their exclusion.

The checks so far compare group medians. Table 12 reports a set of panel specifications that do: each outcome is regressed on an indicator for the battery firms together with firm age, size measured by log revenue and log assets, year fixed effects, and two-digit sector fixed effects, with standard errors clustered by firm.

Table 12. Battery indicator in panel specifications (producing firm-years; controls as described; SE clustered by firm)

| **Outcome** | **All producing years** | **Post-ramp years only** |
|---|---|---|
| Value-added ratio | −0.230*** [−0.375, −0.085] | −0.024 [−0.064, 0.015] |
| Pre-tax margin | −0.282*** [−0.442, −0.121] | −0.073*** [−0.109, −0.037] |
| Log real value added per worker | — | +0.076 [−0.391, 0.542] |

*Notes: Coefficient on the battery indicator, 95% confidence intervals in brackets;***p<0.001. Each outcome is regressed separately on the indicator, firm age, log revenue, log assets, year fixed effects and two-digit sector fixed effects, with standard errors clustered by firm. Post-ramp restricts to firm-years with positive value added; the per-worker specification uses positive observations only. N ranges from 13,950 to 15,035 across 2,054 to 2,230 firms.*

Source: own calculations

Two things follow. The value-added result is instructive in its collapse: the coefficient is large and significant across every specification when all producing years are included, and disappears once ramp years are excluded. The conditional gap in value capture is a construction-phase phenomenon, not a property of mature battery plants — which is what Section 5.6 concluded from the event-study and what the mature-producer specification above showed independently. Three different approaches converge on it.

Profitability behaves differently. The battery firms run a pre-tax margin some seven percentage points below otherwise comparable firms, and this survives the exclusion of ramp years. An obvious objection is that the gap reflects the collapse in European electric-vehicle demand noted above. It does not appear to. Replacing the year and sector fixed effects with sector-by-year effects, which absorb shocks specific to a sector in a given year rather than merely those common to all firms, leaves the coefficient at −6.8 points. Dropping the downturn years altogether gives −5.6 points (excluding 2024) and −6.2 (excluding 2022 onward). Loss-making is neither a construction artefact nor a product of the recent slump: it persists once age, size, sector and year are held constant, and it is present in the years before demand weakened. This result bears directly on H4.

These checks bear on whether the measured differences are artefacts of sample, accounting, or composition. They cannot establish that the strategy caused them. The battery indicator does not vary within firms, so no within-firm identification is available, and the specifications above are conditional comparisons rather than estimates of an effect; Section 5.3 sets out why the design does not attempt more.

## 6. Conclusion

This paper asked what Hungary's battery bet purchased: whether the concentrated exposure it deepened bought movement toward a higher position in the value chain, as a catch-up gamble, or reproduced the country's existing position under a new product, as dependent entrenchment. To answer it I built the exposure–capability framework, which locates a dependent economy's strategic options along the two dimensions a polycrisis has made decisive — the concentration of its external exposure and the depth of its productive capability — and yields four ideal strategies against which a real choice can be placed. I then tested the bet's position against that benchmark with a firm-level panel of Hungary's largest companies over 2010–2024, setting the battery firms beside the foreign-owned export base and the sheltered domestic economy across four hypotheses, and reading the conditions attached to the support that financed them.

The bet expanded enormously. Output, employment and capital multiplied while every comparison group stood still or shrank (H1), confirming the wager is real and large. But expansion alone settles nothing, since it is the precondition both readings share. Where they part, the evidence sides with entrenchment. Value capture does not rise beyond the recovery from construction losses: it matures to the level of comparably young foreign-owned firms, remains below the incumbent export base, and goes no further. Nor does the composition of output shift: seven-eighths of the group's revenue is generated in cell fabrication, the lowest-value stage of the chain, and it stays there (H2, H3). This is the position the dependent-market and downgrading literatures would anticipate (Czirfusz, 2023; Nölke & Vliegenthart, 2009). And the bet does not stand on its own returns: the battery firms are loss-making at the median while expanding fastest, a gap that persists once age, size, sector and year are held constant, and they are sustained by discretionary state support of an intensity no other part of the economy receives (H4). The recent collapse in European demand does not account for this. The gap is present in the years before demand weakened, and it survives specifications that absorb shocks specific to a sector in a given year.

What that support required in return is the finding that the framework most directly delivers. A large subsidy is not in itself evidence of rent; what distinguishes support that builds capability from support that merely sustains volume is the obligations attached to it. The Hungarian scheme's enforceable conditions are volume metrics: investment size, revenue and wage-bill growth, headcount retention, duration. Capability-relevant criteria appear in the decree only as considerations the awarding authority may weigh, never as conditions the recipient must meet, and a capability-contingent instrument that exists within the same scheme was not used for these investments. The state purchased capacity and did not ask for capability. That is a fact about the design of the intervention, established from the statutory record, and it does not depend on how the plants eventually perform.

Together the findings support a single characterisation. The bet is neither an upgrade of the model it joins nor a downgrade of it, but a substitution within it, at constant quality. The economy that once hosted German combustion-engine assembly now hosts Korean and Chinese battery fabrication in the same position: dependent, foreign-owned, and ordinary in what it retains of the value passing through it. What is made has changed without changing where the making sits in the international division of labour. The bet is therefore better read not as a gamble that failed but as one that was never constructed to move the country from where it already stood. The instruments that financed it required volume, and volume is what they bought. Read through the framework, this is what Section 3.3 describes as the entrenchment corner holding — an arrangement reproduced rather than escaped.

Two things bound that conclusion. The first is time. It is a verdict on the bet as observed to 2024, and the plants are young: what the evidence shows is that the sequence which would produce capability has not visibly begun. Were it to begin, it would be visible in the things this design measures — a shift in the composition of output toward materials, components or research functions, and a rising share of value retained as plants mature — and the framework specifies where to look. The second bound concerns how much of the outcome can be attributed to policy at all. Value capture in Poland orders itself by value-chain stage much as it does in Hungary, under a support regime constructed differently, and the region's longest-running cell plant shows the same plateau two years further out. Policy does not appear to determine what a plant at a given stage retains; that seems to follow from the stage itself. What policy can determine is whether production remains at that stage, and here no instrument was directed at moving off it.

The implications run in several directions. For Hungary, the bet does not deliver the escape from dependence its framing promises; it re-tools dependence for the electric age, adding a new concentrated exposure and a heavy subsidy burden. For dependent economies more broadly, the case is a caution: hosting the supply chain of the green transition is not the same as capturing its value, and the scale of an investment reveals nothing about which it delivers. The conditions under which foreign-owned production builds domestic capability are reasonably well established (Section 3.4), and the configuration assembled here meets few of them.

The conditionality finding sharpens this into something a state can act on. The instruments examined here did not lack capability requirements because none were available: the same scheme contained a research-contingent track, with enforceable obligations to create research positions, which was not applied to these investments. Nor is the constraint simply legal — EU rules prohibit conditioning aid on domestic sourcing, but not on research activity, training, or the development of supplier relationships. What distinguishes support that builds capability from support that buys volume is the schedule of obligations attached to it. That schedule is fixed at the point of award, when the state's leverage is greatest and long before any outcome can be observed.

Two further implications follow. The first is a question the spillover literature might take up: its persistent finding of weak or absent linkages in Central and Eastern Europe is usually read as a property of firms or of absorptive capacity, but where the instruments that attracted the investment did not require linkage formation, the absence is at least partly a matter of policy design. The second concerns the assessment of industrial strategy. Outcome-only evaluation cannot distinguish catch-up from entrenchment, because both begin identically. Position and conditionality can be observed within a few years of commitment; capability formation cannot be observed for a generation. A state that has purchased the lowest-value stage of a chain on terms requiring volume and nothing else has already settled much of what follows.

The study has clear limitations. The subsidy that does the most work in the argument is observed from a compiled documented record rather than a firm-year series, and the firm-level support field is structurally blind to the grants that matter most. Foreign ownership is inferred from sector for the comparison groups. Because value capture is derived from statutory accounts, it is exposed to transfer pricing within multinational groups, though the direction of that bias runs against finding upgrading rather than toward it. The panel is also short: most battery firms have produced for fewer than five years, and the largest investment is barely in operation. Most importantly, the analysis measures position and the terms of support, not capability itself: the linkage, spillover and learning channels through which capability accumulates lie outside what firm accounts record, and nothing here establishes that they have failed, only that the movement which would precede them has not occurred. The Polish comparison is confined to the battery chain, without the comparison groups the Hungarian design uses. And the bet is observed in a trough, so a fuller verdict on whether the mature plants can earn their keep must await the market's recovery.

**Declaration of generative AI and AI-assisted technologies in the manuscript preparation process**

During the preparation of this work, the author used Claude (Anthropic) to assemble the firm-level panel from individual data files, to classify and structure the reviewer comments for the revision, to carry out consistency and fact-checks across the manuscript, to refine the structure and style of the text, and to generate the graphic content used in the paper. The author reviewed and edited all output, verified the underlying sources and calculations, and takes full responsibility for the content of the published article.

**Additional material**

[dataset] Crefoport, 2025. Crefoport Scholar firm-level database. https://www.crefoport.hu/scholar

[dataset] Krajowy Rejestr Sądowy, 2025. Repozytorium Dokumentów Finansowych [Repository of Financial Documents]. Ministry of Justice, Poland. https://ekrs.ms.gov.pl (retrieved via https://aleo.com)

[dataset] Eurostat, 2025. Structural business statistics: main indicators by NACE Rev. 2 activity [sbs_ovw_act]. https://ec.europa.eu/eurostat/databrowser/view/sbs_ovw_act

[dataset] Központi Statisztikai Hivatal, 2025. A bruttó hazai termék (GDP) volumen- és árindexei [Volume and price indices of gross domestic product]. Table 21.1.1.1. https://www.ksh.hu/stadat_files/gdp/hu/gdp0001.html

**Data availability**

Data in support of the findings of this study is available from the corresponding author upon reasonable request.

**References**

Acemoglu, D. (2006). A Simple Model of Inefficient Institutions. *The Scandinavian Journal of Economics*, *108*(4), 515–546. https://doi.org/10.1111/j.1467-9442.2006.00460.x

Acemoglu, D., Johnson, S., & Robinson, J. A. (2001). The Colonial Origins of Comparative Development: An Empirical Investigation. *American Economic Review*, *91*(5), 1369–1401. https://doi.org/10.1257/aer.91.5.1369

Aghion, P., & Bircan, C. (2017). *The Middle-Income Trap from a Schumpeterian Perspective* (ADB Economics Working Paper Series No. 521; 0 edn, ADB Economics Working Paper Series). Asian Development Bank. https://doi.org/10.22617/WPS179029-2

Amendolagine, V., Boly, A., Coniglio, N. D., Prota, F., & Seric, A. (2013). FDI and Local Linkages in Developing Countries: Evidence from Sub-Saharan Africa. *World Development*, *50*, 41–56. https://doi.org/10.1016/j.worlddev.2013.05.001

Amendolagine, V., Hansen, U. E., Lema, R., Rabellotti, R., & Ribaudo, D. (2023). Do green foreign direct investments increase the innovative capability of MNE subsidiaries? *World Development*, *170*, 106342. https://doi.org/10.1016/j.worlddev.2023.106342

Amsden, A. H. (2003). *The Rise of the Rest: Challenges to the West from Late-Industrializing Economies*. Oxford University Press, Incorporated. (Original work published 2001)

Antalóczy, K., & Sass, M. (2026). A bizonytalanság kora. *Külgazdaság*, *70*(1–2), 6–17. https://doi.org/10.47630/KULG.2026.70.1-2.6

Bartha, Z. (2024). Changes in varieties of capitalism within the OECD between 2010 and 2020. *International Journal of Public Policy*, *17*(4), 284–301. https://doi.org/10.1504/IJPP.2024.139748

Bartha, Z. (2026). Political influence and corporate profits: A study of Hungarian firms. *Constitutional Political Economy*, *37*(2), 302–323. https://doi.org/10.1007/s10602-026-09505-7

Bartik, T. J. (2020). Using Place-Based Jobs Policies to Help Distressed Communities. *Journal of Economic Perspectives*, *34*(3), 99–127. https://doi.org/10.1257/jep.34.3.99

Bell, M., & Pavitt, K. (1992). Accumulating Technological Capability in Developing Countries. *The World Bank Economic Review*, *6*(suppl 1), 257–281. https://doi.org/10.1093/wber/6.suppl_1.257

Bell, M., & Pavitt, K. (1993). Technological Accumulation and Industrial Growth: Contrasts Between Developed and Developing Countries. *Industrial and Corporate Change*, *2*(2), 157–210. https://doi.org/10.1093/icc/2.2.157

Bhagwati, J. N. (1982). Directly Unproductive, Profit-Seeking (DUP) Activities. *Journal of Political Economy*, *90*(5), 988–1002.

Blackwill, R. D., & Harris, J. M. (2016). *War by Other Means: Geoeconomics and Statecraft*. Harvard University Press. https://doi.org/10.2307/j.ctt1c84cr7

Bohle, D., & Greskovits, B. (2012). *Capitalist Diversity on Europe's Periphery*. Cornell University Press. https://doi.org/10.7591/9780801465666

Bordoff, J., & O'Sullivan, M. L. (2022). Green Upheaval: The New Geopolitics of Energy. *Foreign Affairs*, *101*(1), 130–143.

Cardoso, F. H., & Faletto, E. (1979). *Dependency and Development in Latin America* (M. M. Urquidi, Trans.; 1st edn). University of California Press. https://doi.org/10.2307/jj.13083432

Congleton, R. D., Hillman, A. L., & Konrad, K. A. (Eds). (2008). *40 Years of Research on Rent Seeking 1: Theory of Rent Seeking*. Springer Berlin Heidelberg. https://doi.org/10.1007/978-3-540-79182-9

Council of Ministers of Poland. (2018). *Rozporządzenie Rady Ministrów z dnia 28 sierpnia 2018 r. W sprawie pomocy publicznej udzielanej niektórym przedsiębiorcom na realizację nowych inwestycji [Regulation of 28 August 2018 on State aid granted to certain undertakings for the implementation of new investments]* (Dz.U. 2018 poz. 1713.).

Csaba, L. (2025). Magyarország a világrendetlenségben. *Külgazdaság*, *69*(1–2), 28–35. https://doi.org/10.47630/KULG.2025.69.1-2.28

Czirfusz, M. (2023). *The battery boom in Hungary: Companies of the value chain, outlook for workers and trade unions* (Labour and Social Justice). Friedrich-Ebert-Stiftung.

Dahl, R. A. (1957). The concept of power. *Behavioral Science*, *2*(3), 201–215. https://doi.org/10.1002/bs.3830020303

Damijan, J. P., Knell, M., Majcen, B., & Rojec, M. (2003). The role of FDI, R&D accumulation and trade in transferring technology to transition countries: Evidence from firm panel data for eight transition countries. *Economic Systems*, *27*(2), 189–204. https://doi.org/10.1016/S0939-3625(03)00039-6

Di Bella, G., Flanagan, M., Foda, K., Maslova, S., Pienkowski, A., Stuermer, M., & Toscani, F. (2024). Natural gas in Europe: The potential impact of disruptions to supply. *Energy Economics*, *138*, 107777. https://doi.org/10.1016/j.eneco.2024.107777

Eszterhai, V., & Vörös, Z. (2025, December 1). China vs. Korea: Who Is Winning the Battery Battle in Hungary? *The Diplomat*. https://thediplomat.com/2025/12/china-vs-korea-who-is-winning-the-battery-battle-in-hungary/

European Commission. (2014). *Commission Regulation (EU) No 651/2014 of 17 June 2014 declaring certain categories of aid compatible with the internal market in application of Articles 107 and 108 of the Treaty.* (OJ L 187, 26.6.2014).

European Commission. (2023). *Authorisation for State aid pursuant to Articles 107 and 108 of the Treaty on the Functioning of the European Union—Cases where the Commission raises no objections—SA.63328* (2023/C 195/11). https://eur-lex.europa.eu/legal-content/EN/TXT/?uri=CELEX:52023AS63328

European Commission. (2024a). *Commission Decision (EU) 2024/2057 of 28 February 2023 on the State aid SA.48556 (2019/C) (ex 2018/N)—Hungary—LIP — Regional investment aid to Samsung SDI (notified under document C(2023) 1478)* (OJ L, 2024/2057, 1.8.2024). http://data.europa.eu/eli/dec/2024/2057/oj

European Commission. (2024b). *Commission Implementing Regulation (EU) 2024/2754 of 29 October 2024 imposing a definitive countervailing duty on imports of new battery electric vehicles designed for the transport of persons originating in the People's Republic of China.* (Regulation No. 2024/2754; Official Journal of the European Union L series). European Commission. http://data.europa.eu/eli/reg_impl/2024/2754/oj

Evans, P. B. (1995). *Embedded autonomy: States and industrial transformation*. Princeton University Press.

Fan, H., He, S., & Kwan, Y. K. (2022). FDI forward spillover effects in emerging markets: A comparative meta-analysis of China and Eastern Europe. *Economic Systems*, *46*(3), 101012. https://doi.org/10.1016/j.ecosys.2022.101012

Farrell, H., & Newman, A. L. (2019). Weaponized Interdependence: How Global Economic Networks Shape State Coercion. *International Security*, *44*(1), 42–79. https://doi.org/10.1162/isec_a_00351

Folke, C., Carpenter, S. R., Walker, B., Scheffer, M., Chapin, T., & Rockström, J. (2010). Resilience Thinking: Integrating Resilience, Adaptability and Transformability. *Ecology and Society*, *15*(4), 1–9.

Gagyi, A., & Gerőcs, T. (2025). Dependent development under geopolitical reconfiguration: The Orbán regime in Hungary. *Globalizations*, 1–17. https://doi.org/10.1080/14747731.2025.2491973

Gill, I., & Kharas, H. (2007). *An East Asian Renaissance: Ideas for Economic Growth* [World Bank Publications - Books]. The World Bank Group. https://econpapers.repec.org/bookchap/wbkwbpubs/6798.htm

Government of Hungary. (2014). *210/2014. (VIII. 27.) Korm. Rendelet a beruházás ösztönzési célelőirányzat felhasználásáról [Government Decree 210/2014 (VIII. 27.) on the use of the investment promotion appropriation]* (210/2014 (VIII. 27.)).

Győrffy, D. (2022). The middle-income trap in Central and Eastern Europe in the 2010s: Institutions and divergent growth models. *Comparative European Politics*, *20*(1), 90–113. https://doi.org/10.1057/s41295-021-00264-3

Győrffy, D. (2023). Globalizáció a gazdaságban: Magyarország pozíciója a globális értékláncokban • Economic Globalization: Hungary's Position in Global Value Chains. *Magyar Tudomány*. https://doi.org/10.1556/2065.184.2023.10.6

Hall, P. A., & Soskice, D. (2001). *Varieties of Capitalism: The Institutional Foundations of Comparative Advantage*. Oxford University Press.

Hanousek, J., Kočenda, E., & Maurel, M. (2011). Direct and indirect effects of FDI in emerging European markets: A survey and meta-analysis. *Economic Systems*, *35*(3), 301–322. https://doi.org/10.1016/j.ecosys.2010.11.006

Hauge, J., Houtzager, B., & Hörmann, A. J. (2025). The new economic nationalism: Industrial policy and national security in the United States, China, and the European Union. *Geoforum*, *166*, 104382. https://doi.org/10.1016/j.geoforum.2025.104382

Helleiner, E. (2024). Economic Globalization's Polycrisis. *International Studies Quarterly*, *68*(2), sqae024. https://doi.org/10.1093/isq/sqae024

Hidalgo, C. A., Klinger, B., Barabási, A.-L., & Hausmann, R. (2007). The Product Space Conditions the Development of Nations. *Science*, *317*(5837), 482–487. https://doi.org/10.1126/science.1144581

Hirschman, A. O. (1945). *National Power and the Structure of Foreign Trade* (1st edn). University of California Press. https://doi.org/10.2307/jj.15976659

Javorcik, B. S., & Spatareanu, M. (2008). To share or not to share: Does local participation matter for spillovers from foreign direct investment? *Journal of Development Economics*, *85*(1–2), 194–217. https://doi.org/10.1016/j.jdeveco.2006.08.005

Juhász, R., Lane, N., & Rodrik, D. (2024). The New Economics of Industrial Policy. *Annual Review of Economics*, *16*(1), 213–242. https://doi.org/10.1146/annurev-economics-081023-024638

Keohane, R. O., & Nye, J. S. (2012). *Power and interdependence* (4. ed). Pearson.

Krueger, A. O. (1974). The Political Economy of the Rent-Seeking Society. *American Economic Review*, *64*(3), 291–303.

Kuik, C.-C. (2021). Getting hedging right: A small-state perspective. *China International Strategy Review*, *3*(2), 300–315. https://doi.org/10.1007/s42533-021-00089-5
Lall, S. (1992). Technological capabilities and industrialization. *World Development*, *20*(2), 165–186. https://doi.org/10.1016/0305-750X(92)90097-F
Lawrence, M., Homer-Dixon, T., Janzwood, S., Rockström, J., Renn, O., & Donges, J. F. (2024). Global polycrisis: The causal mechanisms of crisis entanglement. *Global Sustainability*, *7*, e6. https://doi.org/10.1017/sus.2024.1
Lewicka, E. D., Burkowicz, A., Babii, K., Riumina, D., Szlugaj, J., Kamyk, J., & Gałaś, A. (2025). The impact of the war in Ukraine on the supply of selected raw materials to Poland and to the European Union. *Gospodarka Surowcami Mineralnymi - Mineral Resources Management*, 5–43. https://doi.org/10.24425/gsm.2025.155341
Malak, M., Panton, A., Rojas-Romagosa, H., & Weller, A. (2025). Promoting Energy Security in Hungary: A Model-based Analysis. *Selected Issues Papers*, *2025*(121), 1–16. https://doi.org/10.5089/9798229023177.018
Miller, C. (2022). *Chip War: The fight for the world's most critical technology*. Simon & Schuster.
Morin, E., & Kern, A. B. (1999). *Homeland earth: A manifesto for the new millennium*. Hampton Press. (Original work published 1993)
Muraközy, B., Bisztray, M., & Reizer, B. (2018). *Productivity differences in Hungary and mechanisms of TFP growth slowdown.* Publications Office of the European Union. https://doi.org/10.2873/33213
Murphy, K. M., Shleifer, A., & Vishny, R. W. (1993). Why Is Rent-Seeking So Costly to Growth? *The American Economic Review*, *83*(2), 409–414.
Naseemullah, A. (2023). The political economy of national development: A research agenda after neoliberal reform? *World Development*, *168*, 106269. https://doi.org/10.1016/j.worlddev.2023.106269
Nölke, A., & Vliegenthart, A. (2009). Enlarging the Varieties of Capitalism: The Emergence of Dependent Market Economies in East Central Europe. *World Politics*, *61*(4), 670–702. https://doi.org/10.1017/S0043887109990098
Pavlinek, P., Allez-Aller, R., Gil-Canaleta, C., & Ullibarri-Arce, M. (2017). *Foreign Direct Investment and the Development of the Automotive Industry in Eastern and Southern Europe* (2017.03). European Trade Union Institute. https://www.etui.org/sites/default/files/17%20WP%202017%2003%20Foreign%20Direct%20Investment%20and%20the%20development%20EN%20Web%20version.pdf
Ricz, J., & Éltető, A. (2025). Dependency meets illiberalism: Expansion of the EV battery sector in Central Europe. *Post-Communist Economies*, 1–26. https://doi.org/10.1080/14631377.2025.2587008
Rodrik, D. (2012). *The globalization paradox: Democracy and the future of the world economy* (Norton paperback). W. W. Norton & Company.
Rodrik, D., Grossman, G., & Norman, V. (1995). Getting Interventions Right: How South Korea and Taiwan Grew Rich. *Economic Policy*, *10*(20), 53. https://doi.org/10.2307/1344538
Rumelt, R. P. (2017). *Good strategy, bad strategy: The difference and why it matters*. Profile Books.
Scheiring, G. (2021). Dependent development and authoritarian state capitalism: Democratic backsliding and the rise of the accumulative state in Hungary. *Geoforum*, *124*, 267–278. https://doi.org/10.1016/j.geoforum.2019.08.011
Shkolnykova, M. (2026). The impact of FDI-related inventor mobility on the innovation performance of domestic firms: Evidence from the V4 countries. *Economic Systems*, *50*(1), 101349. https://doi.org/10.1016/j.ecosys.2025.101349
Solow, R. M. (1957). Technical Change and the Aggregate Production Function. *The Review of Economics and Statistics*, *39*(3), 312. https://doi.org/10.2307/1926047
State Audit Office of Hungary. (2025). *Külföldi működőtőke beruházásokhoz nyújtott közvetlen költségvetési támogatások szempontrendszere, és azok érvényesülése [Criteria for direct budgetary support to foreign direct investment, and their implementation]* (Analysis EL-4006-004/2025). State Audit Office of Hungary. https://www.asz.hu/dokumentumok/EKD_elemzes_nyilvanossagra_0422.pdf

Subramanian, A., & Kessler, M. (2014). The Hyperglobalization of Trade and Its Future. In F. Allen, J. R. Behrman, N. Birdsall, S. Fardoust, D. Rodrik, A. Steer, & A. Subramanian, *Towards a Better Global Economy* (1st edn, pp. 216–288). Oxford University PressOxford. https://doi.org/10.1093/acprof:oso/9780198723455.003.0004

Tagliapietra, S., Trasi, C., & Sebastian, G. (2025). *A smart European strategy for electric vehicle investment from China* (No. 21/2025). Bruegel. https://www.bruegel.org/policy-brief/smart-european-strategy-electric-vehicle-investment-china

Tooze, A. (2022, October). Welcome to the World of Polycrisis. *Financial Times*. https://www.ft.com/content/498398e7-11b1-494b-9cd3-6d669dc3de33

Transport & Environment. (2025). *Assembly plant or battery powerhouse? Analysis of foreign battery investments in EU*. Transport & Environment. https://www.transportenvironment.org/uploads/files/Assembly-plant-or-battery-powerhouse.pdf

Tullock, G. (1967). The Welfare Cost of Tariffs, Monopolies, and Theft. *Economic Inquiry*, *5*(3), 224–232. https://doi.org/10.1111/j.1465-7295.1967.tb01923.x

Ustawa z Dnia 10 Maja 2018 r. o Wspieraniu Nowych Inwestycji [Act of May 10, 2018 on Support for New Investments], Poz. 1162 (2018). https://api.sejm.gov.pl/eli/acts/DU/2018/1162/text.pdf

Voszka, É., Végh, M. Z., & London, A. (2023). Támogatási nagyüzem – az egyedi kormánydöntések példája. *Közgazdasági Szemle*, *70*(12), 1297–1337. https://doi.org/10.18414/KSZ.2023.12.1297

Walker, B., Holling, C. S., Carpenter, S., & Kinzig, A. (2004). Ecology and Society: Resilience, Adaptability and Transformability in Social–ecological Systems. *Ecology and Society*, *9*(2), 1–9. https://doi.org/10.5751/ES-00650-090205

Walt, S. M. (1987). *The Origins of Alliances*. Cornell University Press. https://doi.org/10.7591/9780801469992

World Bank. (2024). *World Development Report 2024: The Middle-Income Trap*. Washington, DC: World Bank. https://doi.org/10.1596/978-1-4648-2078-6